\documentclass[aps,prb,twocolumn,groupedaddress,showpacs,superscriptaddress,amssymb,amsmath]{revtex4-1}
\usepackage{amsmath}
\usepackage{latexsym}
\usepackage{amssymb}
\usepackage{graphics,epstopdf}
\usepackage{hyperref}
\usepackage{soul}
\usepackage{xcolor}
\usepackage{natbib}
\hypersetup{
    colorlinks = true,
    linkcolor =blue,
	citecolor=blue, 
	urlcolor=blue 
}

\newcommand{\ed}{\end{document}}
\newcommand{\beq}{\begin{equation}}
\newcommand{\eeq}{\end{equation}}

\usepackage{mathtools}
\usepackage{comment}

\begin{document}
%\title{Consequences of non-Hermitian transfer matrix properties in open-system conductance of finite-range hopping model}

%\title{Consequence of non-Hermitian transfer matrices and exceptional hyper-surfaces in non-equilibrium steady state conductance of finite-range lattice models}

%\title{Consequence of non-Hermitian transfer matrices and exceptional hyper-surfaces in %quantum transport properties of finite-range lattice models}

\title{Exceptional hypersurfaces of transfer matrices of finite-range lattice models and their consequences on quantum transport properties}
\author{Madhumita Saha}
\email{madhumita.saha@icts.res.in }
\affiliation{Department of Physics, Indian Institute of Science Education and Research Pune, Dr. Homi Bhabha Road, Ward No. 8, NCL Colony, Pashan, Pune, Maharashtra 411008, India} 
\affiliation{International Centre for Theoretical Sciences, Tata Institute of Fundamental Research,
Bangalore 560089, India}

\author{Manas Kulkarni}
\email{manas.kulkarni@icts.res.in} 
\affiliation{International Centre for Theoretical Sciences, Tata Institute of Fundamental Research,
Bangalore 560089, India}

\author{Bijay Kumar Agarwalla}
\email{bijay@iiserpune.ac.in}
\affiliation{Department of Physics, Indian Institute of Science Education and Research Pune, Dr. Homi Bhabha Road, Ward No. 8, NCL Colony, Pashan, Pune, Maharashtra 411008, India}

\begin{abstract}
We investigate the emergence and corresponding nature of exceptional points located on exceptional hyper-surfaces of non-Hermitian transfer matrices for finite-range one-dimensional lattice models. We unravel the non-trivial role of these exceptional points in determining the system size scaling of electrical conductance in non-equilibrium steady state. We observe that the band edges of the system 
always correspond to the transfer matrix exceptional points. Interestingly, albeit the lower band edge always occurs at wave-vector $k=0$, the upper band edge may or may not correspond to $k=\pi$. Nonetheless, in all the cases, the system exhibits universal subdiffusive transport
for conductance at every band edge with scaling $N^{-b}$ with scaling exponent $b= 2$. However, for cases when the upper band edge is not located at $k=\pi$, 
the conductance  features interesting oscillations with overall $N^{-2}$ scaling. Our work further reveals that this setup is uniquely suited to 
%systematically 
generate higher order transfer matrix exceptional points at upper band edge when one considers finite range hoppings beyond nearest neighbour. Additional exceptional points other than those at band edges are shown to occur, although interestingly, these do not give rise to anomalous transport.  

%Beyond nearest neighbour hopping model, we find that additional transfer matrix exceptional point is possible (within band edge or outside the band edge) when the chemical potential is not at the band edge. But, these exceptional points do not give rise to universal subdiffusive transport. 

%We also find that odd and even number of hoppings can show different behaviour in conductance at the upper band edge depending on the conditions on hoppings.

\end{abstract}
\date{\today}
\maketitle

%\tableofcontents

\section{Introduction}
Understanding of the emergence of exceptional points and exceptional surfaces in non-Hermitian Hamiltonian systems is an active and rapidly growing area of research \cite{exceptional_surface1,exceptional_surface2,exceptional_surface3,exception_surface4,higher_order_exceptional,toplology_non_Hermitian,sensing1,sensing_review,sensing_review1}. Typically these exceptional points are extremely sensitive to external perturbations and therefore are useful for potential applications in cavity quantum electrodynamics, spectral filtering, sensing, lasing, and thermal imaging \cite{exception_surface4,sensing1,sensing_review,sensing_review1}. Moreover, exceptional hyper-surfaces i.e., hyper-surfaces hosting exceptional points, are more beneficial than a discrete exceptional point. This is because, in realistic setups, tuning and stabilizing a system to a discrete exceptional point, especially in large parameter space is highly challenging and often impossible \cite{exceptional_surface1,exceptional_surface2,exceptional_surface3,exception_surface4}. Similar to non-Hermitian Hamiltonian,  for one-dimensional (1D) nearest neighbour tight-binding systems, the underlying non-Hermitian transfer matrix of the lattice is known to have exceptional points at the band edges \cite{Madhumita_subdiffusive}. However exceptional hyper-surfaces (i.e., higher dimensional) of transfer matrices for such lattice systems have not been reported earlier. Interestingly, beyond the nearest neighbor hopping model, due to the increased dimensionality of the transfer matrix, there is a strong possibility of the emergence of exceptional hyper-surfaces and thereby higher-order exceptional points. One of the main aims of this work is to unravel the nature of transfer matrices for finite-range hopping model (involving $n$ number of neighbors where $n$ does not scale with system size $N$).

Understanding non-equilibrium steady-state transport properties in low-dimensional lattice systems is another important area of research \cite{low_dimensional1,low-dimensional2,low-dimensional3,low-dimensional4,low-dimensional5,low-dimensional6}. This is crucial both from a fundamental perspective as well as from a technological point of view. A deep understanding of transport behaviour is paramount to realise efficient quantum  devices \cite{device1,device2,device3,buttikerdephase}. The study of quantum transport in low-dimensional systems is interesting as often it shows deviation from the normal diffusive behaviour or standard Ohm's law/Fourier's law which is one of the reasons why low-dimensional systems have been of fundamental interest \cite{low_dimensional1,low-dimensional2,low-dimensional3}. Sample examples of low-dimensional systems include 1D, 2D systems with random and quasi-periodic disorder \cite{Andersion_random,2D_Andersion_model,Pandit83,disordered_electronic,Aubry,Correlated_disorder,Commensurate_AAH,GAAH_mobility_edge,
Archak_AAH,Archak_AAH1,Quasi-periodic_cecilia}.
For the random disorder case, Anderson localization occurs in 1D and 2D which essentially unravels the exponential nature of localization of all the single particle states \cite{Andersion_random,Pandit83,2D_Andersion_model}. As a consequence of exponentially localized single particle states, the transport exponentially decays as a function of system size $N$. Recall that, in absence of disorder, transport is independent of system size (ballistic transport) \cite{low_dimensional1,low-dimensional5}. Therefore, both the clean and disordered systems show deviations from normal diffusive behaviour akin to the Ohm's law.

Quasi-periodic disordered systems in low dimensions are known to show unusual and rich transport properties. The study of transport properties in low-dimensional quasi-periodic systems gained a lot of attention because of very successful experimental realizations in various platforms \cite{I_bloch_experiment,expt1,expt2,expt3,GAAH_experiment,AAH-expt-Silberberg,zilberberg1,zilberberg2}. These systems often show anomalous transport in different cases \cite{Archak_AAH,Archak_phase_diagram,Archak_AAH1,Quasi-periodic_cecilia,anomalous1,anomalous2,anomalous3}. Note that, in anomalous transport conductance $\mathcal{G}\sim N^{-b}$ where $0<b \neq 1$. $b=1$ is the limit of diffusive transport. $b>1$ refers to subdiffusive transport whereas $0<b<1$ refers to  superdiffusive transport. Though such quasi-periodic disordered systems often show anomalous transport, the microscopic understanding of the anomalous transport is far from being fully understood.
%{\color{red}{stop}}

Interestingly, in a recent work, it was shown that for a 1D nearest neighbor tight-binding fermionic lattice with periodic on-site potential, the conductance displays subdiffusive scaling at the band edges of the system.
The origin of this effect was shown to be connected to the presence of exceptional points corresponding to the non-Hermitian transfer matrices of the lattice \cite{Madhumita_subdiffusive}. Moreover, such a subdiffusive scaling at the band edges was also observed in long-range lattice systems with power-law hopping (involving $n$ numbers of neighbors where $n$ scales with system size $N$) \cite{Archak_long_range} where the transfer matrix approach is not well suited. The understanding behind this effect for long-range systems is still lacking. To bridge the gap between nearest neighbor hopping systems and long-range hopping systems, investigation of the transport properties of finite-range hopping model and its connection with underlying transfer matrices is crucial.

%In this work, we aim to bridge the gap between nearest neighbor hopping systems and long-range hopping systems by considering finite-range hopping model (involving $n$ number of neighbors where $n$ does not scale with system size). 

In this work, we provide an in-depth understanding of emergence of exceptional hyper-surfaces of transfer matrices in finite-range hopping model  and their impact on non-equilibrium-steady state (NESS) quantum transport properties. Our main findings can be summarized as follows,

\begin{enumerate}
    \item We establish the non-trivial connection between non-equilibrium steady state (NESS) conductance and underlying $2n\times 2n$ dimensional non-Hermitian transfer matrix for finite-range lattice models.

    \item We always find the appearance of exceptional points of various orders at the band edges of the lattice, which crucially depends on $n$. It is important to note that by ``points", we also mean ``hyper-surfaces" in more general sense. We unravel the non-trivial role played 
by these exceptional points in determining the universal system size scaling of NESS conductance with scaling exponent $b=2$. We further demonstrate that the value of the scaling exponent is remarkably robust to the order of the exceptional point.

\item We find that for finite-range hopping model ($n>1$) the location of the upper band edge does not always corresponds to $k=\pi$. In such cases, we observed interesting oscillation features in conductance with overall $N^{-2}$ scaling.  
\end{enumerate}

The plan of the paper is as follows: In section \ref{sec:LH_DR}, we provide the  lattice Hamiltonian and dispersion relation for the finite-range hopping model. In section \ref{sec:OQS_properties}, we discuss the open-system transport properties. First, we provide the non-equilibrium-steady-state (NESS) conductance in detail (section.~\ref{subsec:NESS_conductance}). To calculate conductance, its important to compute the Green's function. It can be calculated using transfer matrix approach. Thus, in section.~\ref{conn-rt} we discuss the connection between transfer matrices and NESS conductance. This connection involves the exponents of transfer matrices. In section.~\ref{sec:TM}, we discuss the details of eigenvalues (section.~\ref{subsec:eigenvalue}), eigenvectors (section.~\ref{subsec:eigenvector}) and exponents of transfer matrices (section.~\ref{subsec:exponents}). In section.~\ref{sec:result}, we first provide the results for transfer matrix properties for finite-range hopping model with $n=2$ followed by the scaling of NESS conductance (section.~\ref{n-2-sec}) . Next, in section.~\ref{general-sec} we provide generalizations of some results beyond $n=2$. Then, in section.~\ref{robust} we discuss the robustness of some results. Finally, in section.~\ref{sec:summary}, we conclude and discuss future directions. In Appendix.~\ref{sec:appA} we provide detailed calculations establishing the connection between transfer matrix and conductance. 
In Appendix.~\ref{sec:app2}, we show the nature of transfer matrix eigenvalues analytically for $n=2$.

%%%%%%%%%%%%%%%%%%%%%%%%%%%%%%%%%%%%%%%%%%%%%%%%%%%%%%%%%%%
%%%%%%%%%%%%%%%%%%%%%%%%%%%%%%%%%%%%%%%%%%%%%%%%%%%%%%%%%%%
%%%%%%%%%%%%%%%%%%%%%%%%%%%%%%%%%%%%%%%%%%%%%%%%%%%%%%%%%%%
%%%%%%%%%%%%%%%%%%%%%%%%%%%%%%%%%%%%%%%%%%%%%%%%%%%%%%%%%%%
%%%%%%%%%%%%%%%%%%%%%%%%%%%%%%%%%%%%%%%%%%%%%%%%%%%%%%%%%%%
%%%%%%%%%%%%%%%%%%%%%%%%%%%%%%%%%%%%%%%%%%%%%%%%%%%%%%%%%%%
%%%%%%%%%%%%%%%%%%%%%%%%%%%%%%%%%%%%%%%%%%%%%%%%%%%%%%%%%%%
\section{Lattice Hamiltonian and Dispersion Relation}
\label{sec:LH_DR}

% In this section, first in subsection A, we describe the tight-binding Hamiltonian and the dispersion for finite-range hopping model. As we are interested in the non-equilibrium steady state current for the finite-range hopping model, we discuss the open system conductance using non-equilibrium Green's function (NEGF) formalism in subsection B. The Green's function of the finite-range hopping model is nothing but the inversion of symmetric banded Toeplitz matrix. To calculate that we need the transfer matrices. Thus, the connection between Green's function and transfer matrix is given in subsection C. To calculate the Green's function, it is also important to understand the transfer matrix eigenvalues, eigenvectors and the matrix elements of exponents of transfer matrix. We discuss these things in subsection D, E and F respectively.

% \subsection{Model Hamiltonian and Dispersion}
In this section, we introduce the tight-binding Hamiltonian for the finite-range hopping model and provide the details of the dispersion relation. The Hamiltonian for our set-up is given as,
\begin{align}
\label{model}
\hat{H}=-\sum\limits_{i=1}^{N}\sum\limits_{m=1}^{n} t_{m} \hat{c}_{i}^{\dagger} \hat{c}_{i+m} + \mathrm{h.c.} 
\end{align}
Here $\hat{c}_{i}^{\dagger}(\hat{c}_i)$ is the fermionic creation (annihilation) operator. $t_{m}$ is the hopping strength for $m-$th neighbor. We consider the lattice size as $N$ with $n$ being the total number of neighbors to the left and to the right of a particular lattice site, if any. In the thermodynamic limit, the dispersion relation for this set-up is given as,
\begin{align}
\label{dispersion}
\omega(k)=-2\sum\limits_{m=1}^n t_m \cos{mk}.
\end{align}
We immediately notice from Eq.~\ref{dispersion} that the minimum value of $\omega(k)$ which corresponds to the lower band edge always occurs at wave-vector value $k=0$. The value of the energy at the lower band edge ($k=0$) is,
\begin{align}
\label{lower band edge}
\omega(k=0)=-2\sum\limits_{m=1}^n t_m.
\end{align}
Interestingly, the maximum value of $\omega(k)$ which corresponds to upper band edge may or may not always occur at $k=\pi$ and crucially depends on the range of hopping $n$ and strength of hopping $t_m$. We now find a condition which decides whether or not $k=\pi$ is an upper band edge. For that purpose, we use Eq.~\ref{dispersion} and demand (negative second derivative implying maxima at $k=\pi$), 
%\begin{align}
%\label{maxcond}
%\frac{d^2 \omega(k)}{dk^2}=  2 \sum \limits_{m=1}^n m^2 t_m \cos mk.
%\end{align}  
\begin{align}
\label{maxcond1}
\frac{d^2 \omega(k)}{dk^2}\Big{|}_{k=\pi}=  2 \sum \limits_{m=1}^n (-1)^m m^2 t_m < 0 .
\end{align}
Note that, at the extremum (maxima or minima) of $\omega(k)$, the first derivative of $\omega(k)$ always vanishes. Thus, from Eq.~\ref{maxcond1} we immediately receive the condition for getting the upper band edge at $k=\pi$ which is given as,
\begin{align}
\label{condition1}
\sum\limits_{m \in \rm{even}}m^2 t_m \! < \!\sum\limits_{m \in \rm{odd}}m^2 t_m.
\end{align}
Note that, the sum in above Eq.~\ref{condition1} is over the range of hopping only.
 With this condition (Eq.~\ref{condition1}) being satisfied 
and using Eq.~\ref{dispersion}, we see that the value for upper band edge energy is given as,  
\begin{align}
\label{upper band edge}
\omega(k=\pi)=2 \sum\limits_{m=1}^n (-1)^{m+1} t_m.
\end{align}
Now, for the condition, 
\begin{align}
\label{condition2}
\sum\limits_{m\in \rm{even}}m^2 t_m > \sum\limits_{m \in \rm{odd}}m^2 t_m
\end{align}
the upper band edge will occur at some different $k=k_1\neq \pi$.

An interesting situation appears when, 
\begin{align}
\label{condition3}
\sum\limits_{m\in \rm{even}}m^2 t_m = \sum\limits_{m \in \rm{odd}}m^2 t_m
\end{align}
in which case the second derivative in Eq.~\ref{maxcond1} disappears. Hence one needs to look at the higher order derivatives to conclude about the upper band edge. Now, as the third derivative at $k=\pi$ is always zero, we look at the fourth order derivative which is given as,
\begin{align}
\label{maxcond2}
\frac{d^4 \omega(k)}{dk^4}\Big{|}_{k=\pi}=  2 \sum \limits_{m=1}^n (-1)^{m+1} m^4 t_m  .
\end{align}

Interestingly, for even number of hoppings ($n$=even) along with the condition in Eq.~\ref{maxcond2}, we find that
\begin{equation}
  \frac{d^4 \omega(k)}{dk^4}\Big{|}_{k=\pi} <0   
\end{equation}
which implies a local maximum at $k=\pi$. Thus, in such a scenario the $\omega(k)$ value given in Eq.~\ref{upper band edge} is the upper band edge. However for odd number of hoppings ($n$=odd), the condition in Eq.~\ref{maxcond2} implies 
\begin{equation}
\frac{d^4 \omega(k)}{dk^4}\Big{|}_{k=\pi}>0    
\label{high-der}
\end{equation}
is greater than zero, ensuring a local minimum. But, since $k=0$ always corresponds to a global minimum (lower band edge), in this situation $k=\pi$ will not correspond to any band edge and  therefore the upper band edge will occur at some different $k=k_1$ where $k_1\neq \pi$.

 In summary, the above analysis points out that for finite-range hopping model, $k=0$ always corresponds to the lower band edge. But, $k=\pi$ may or may not correspond to the upper band edge and depends crucially on the conditions as given in Eqs.~\ref{condition1},\ref{condition2}, \ref{condition3}.  In what follows, we will see interesting consequences of this fact in the NESS transport properties. It is important to note that for the nearest neighbor hopping model i.e., $n=1$, the upper band edge is always at $k=\pi$ which is also clear from Eq.~\ref{dispersion} and Eq.~\ref{maxcond1}.
 %{\color{red}{STOP}}
 %With even number of hoppings, the condition for $k=\pi$ correspond to the upper band edge is $\sum\limits_{m\in even}m^2 t_m \! \leq \!\sum\limits_{m\in odd}m^2 t_m$ and with odd number of hoppings, the condition for $k=\pi$ correspond to the upper band edge is $\sum\limits_{m \in even}m^2 t_m \! < \!\sum\limits_{m \in odd}m^2 t_m$. Thus for general finite-range hopping model, depending on the condition that $k=\pi$ corresponds to the upper band edge or not, we have three scenarios (a) $\sum\limits_{m \in even}m^2 t_m \! < \!\sum\limits_{m \in odd}m^2 t_m$ (b)$\sum\limits_{m \in even}m^2 t_m \! > \!\sum\limits_{m \in odd}m^2 t_m$ and (c) $\sum\limits_{m \in even}m^2 t_m \! = \!\sum\limits_{m \in odd}m^2 t_m$. 

%%%%%%%%%%%%%%%%%%%%%%%%%%%%%%%%%%%%%%%%%%%%%%%%%%%%%%%%%%%
%%%%%%%%%%%%%%%%%%%%%%%%%%%%%%%%%%%%%%%%%%%%%%%%%%%%%%%%%%%
\section{Open quantum system transport properties}
\label{sec:OQS_properties}
\subsection{Non-equilibrium steady state conductance}
\label{subsec:NESS_conductance}
In this section, we are interested in computing the NESS conductance when the finite range hopping lattice chain is connected with two fermionic baths at its two ends i.e., at site $1$ and site $N$. The baths are modelled by infinite number of fermionic modes and the associated spectral functions are denoted by $\mathcal{J}_1(\omega)$ and $\mathcal{J}_N(\omega)$, respectively. At the initial time $t=0$, both the baths are kept at zero temperature ($\beta=\infty$) but at slightly different chemical potentials $\mu$ and $\mu-\delta \mu$, respectively. The finite lattice system can however be in any arbitrary initial state. Note that, if the bandwidth of the baths is larger than the bandwidth of the system, the lattice system usually reaches a unique NESS in the long-time limit.

In this study, we are interested in the linear response regime and NESS conductance. 
%therefore consider $\delta \mu$ as small. 
Using the non-equilibrium-Green's-Function (NEGF) \cite{keldysh1,keldysh2,keldysh3,keldysh4,Wang2014,Wingreen}, we can write down the NESS conductance as \cite{Wingreen}
\begin{align}
\label{conductance}
\mathcal{G}(\mu)=\frac{1}{2\pi}\mathcal{J}_1(\mu) \mathcal{J}_N (\mu) |\mathbf{G}_{1N}(\mu)|^2.
\end{align}
Here $\mathbf{G}(\mu)$ is the $N \times N$ retarded NEGF matrix and is given as 
\begin{equation}
\label{negf}
\mathbf{G}(\mu)=\Big[\mu \, \mathbb{I}-\mathbf{H}-\mathbf{\Sigma}_{1}-\mathbf{\Sigma}_N\Big]^{-1},
\end{equation} 
where $\mathbf{H}$ is the $N \times N$ single-particle lattice Hamiltonian matrix corresponding to $\hat{H}$ in Eq.~\ref{model}, $\mathbb{I}$ is a $N\times N$ identity matrix, $\mathbf{\Sigma}_{1}$ and $\mathbf{\Sigma}_N$ are the diagonal $N \times N$ self-energy matrices for the left and right baths with non-zero entries only at $\big(\mathbf{\Sigma}_{1}\big)_{11}$  and $\big(\mathbf{\Sigma}_{N}\big)_{NN}$.
%which are attached at $1$st and $N$th sites of the system.  
%Open-system transport properties can be classified in terms of the system size scaling of conductance. Now, for the diffusive transport conductivity is system size independent which is the usual normal behaviour. In this case, conductance $\mathcal{G}(\mu)\sim 1/N$. But, in low-dimensional systems, we often see deviation from this behaviour. For low-dimensional transport, two extreme cases are (1) ballistic transport $\mathcal{G}(\mu)\sim N^0$ (2) `no transport' $\mathcal{G}(\mu)\sim e^{-N/\xi}$, where $\xi $ is the localization length. In between, anomalous transport properties can be also seen in low-dimensional systems where $\mathcal{G}\sim N^{-\delta}$ with $0<\delta \neq 1$. Here, when $0<\delta<1$, this is known as superdiffusive transport and with $\delta>1$, it is known as subdiffusive transport. 
Thus, following Eq.~\ref{conductance}, to infer the scaling property of conductance $\mathcal{G}(\mu)$ with system size $N$ we need to investigate the system size scaling only for $G_{1N}(\mu)$ as the spectral functions, being the property of the baths, are independent of $N$. Moreover, as the baths are attached at the two ends of the lattice, the scaling of $\mathbf{G}_{1N}(\mu)$ with $N$ is directly governed by the scaling of the bare part of the retarded Green's function $\mathbf{g}_{1N}(\mu)$ \cite{Archak_long_range} defined as 
\begin{equation}
\label{bare-GF}
\mathbf{g}(\mu)=\big[\mu \,\mathbb{I} - \mathbf{H}\big]^{-1}.
\end{equation}
In section \ref{conn-rt}, we focus on the calculation of $\mathbf{g}(\mu)$ for the finite-range model by introducing the transfer matrix approach. 

%we can infer about the transport properties from the system size scaling of $|G_{1N}(\mu)|^2$ (from Eq.~\ref{conductance}) as well as bare Green's function $|g_{1N}(\mu)|^2$. The bare Green's function $g(\mu)=(\mu \mathbb{I}-\mathbf{H})^{-1}$. In the next section, we focus on the calculation of bare Green's function for the finite range hopping model.

%$|G_{1N}|^2$ with system size $N$. As, the self energy terms does not have any $N$ dependence, we can conclude about the system size scaling from bare Green's function ($|g_{1N}(\omega)|^2$) also. Here $g(\omega)=(\omega \mathbb{I}-\mathbf{H_s})^{-1}$.

%%%%%%%%%%%%%%%%%%%%%%%%%%%%%%%%%%%%%%%%%%%%%%%%%%%%%%%%%%%
%%%%%%%%%%%%%%%%%%%%%%%%%%%%%%%%%%%%%%%%%%%%%%%%%%%%%%%%%%%
%%%%%%%%%%%%%%%%%%%%%%%%%%%%%%%%%%%%%%%%%%%%%%%%%%%%%%%%%%%
%%%%%%%%%%%%%%%%%%%%%%%%%%%%%%%%%%%%%%%%%%%%%%%%%%%%%%%%%%%

\subsection{Connection between retarded bare Green's function and transfer matrix}
\label{conn-rt}
%For the finite range hopping model, the bare Green's function is,
%\begin{align}
%\mathbf{g}(\mu)=(\mu \mathbb{I}-\mathbf{H})^{-1}.
%\end{align}
In this section to facilitate further discussion, we first provide the details of the transfer matrix for the finite-range lattice model with hopping range $n$ \cite{transfer_matrix1,transfer_matrix2}. To construct the transfer matrix, we write the discrete version of the time-independent Schr{\"o}dinger equation $\hat{H} |\psi\rangle= \omega |\psi\rangle$ as,

\begin{align}
\label{discretescro}
\omega \psi_{\ell}= -t_1 \psi_{\ell+1} - t_1 \psi_{\ell-1}- t_2 \psi_{\ell+2}- t_2 \psi_{\ell-2}+ \ldots \\ \nonumber
- t_n \psi_{\ell+n} - t_n \psi_{\ell-n},
\end{align}

where $\psi_{\ell}$ is the amplitude of wave-function at the $\ell$th site.  We can rewrite Eq.~\ref{discretescro} as,
\begin{align}
\label{discretescro2}
\psi_{\ell+n}=-\frac{\omega}{t_n} \psi_{\ell} -\frac{t_1}{t_n} \Big{[}\psi_{\ell-1}+\psi_{\ell+1} \Big{]} - \\ \nonumber    
\frac{t_2}{t_n} \Big{[}\psi_{\ell-2}+\psi_{\ell+2} \Big{]} + \ldots - \psi_{\ell-n}
\end{align}
Following Eq.~\ref{discretescro2}, we can write how the amplitude of wave-function at $(\ell+n)$, $(\ell+n-1)$, $\ldots (\ell-n+2)$, $(\ell-n+1)$th sites are connected with $(\ell+n-1)$,$(\ell+n-2) \ldots$, $(\ell-n+1)$, $(\ell-n)$  th site via a $2n \times 2n$ transfer matrix $\mathbf{T}^{(l)}(\omega)$, given as, 
\begin{widetext}
\begin{align}
\label{discreteschro3}
\begin{pmatrix} \psi_{\ell+n} \\ \psi_{\ell+n-1}\\ \vdots \\ \psi_{\ell-n+3}\\ \psi_{\ell-n+2}\\ \psi_{\ell-n+1}   \end{pmatrix}&= \begin{pmatrix} -\frac{t_{n-1}}{t_n} & -\frac{t_{n-2}}{t_n}& \ldots & -\frac{\omega}{t_n} & \ldots & -\frac{t_{n-2}}{t_n} & -\frac{t_{n-1}}{t_n} & -1 \\
1 & 0 & 0 & \ldots & \ldots & 0 & 0 &0 \\
0 & 1 & \ldots & 0 & \ldots & 0 & 0 &0 \\
\vdots & \vdots & \vdots & \vdots & \vdots & \vdots & \vdots & \vdots \\
0 & 0 & \ldots & \ldots & \ldots & \ldots & 0 & 0\\
0 & 0 & \ldots & \ldots & \ldots & \ldots & 1 & 0
\end{pmatrix}  
\begin{pmatrix} \psi_{\ell+n-1} \\ \psi_{\ell+n-2}\\ \vdots \\ \psi_{\ell-n+2}\\ \psi_{\ell-n+1} \\ \psi_{\ell-n}   \end{pmatrix} 
= \mathbf{T}^{(\ell)}(\omega) \begin{pmatrix} \psi_{\ell+n-1} \\ \psi_{\ell+n-2}\\ \vdots\\ \psi_{\ell-n+2} \\ \psi_{\ell-n+1} \\ \psi_{\ell-n}   \end{pmatrix} . 
\end{align}
\end{widetext}
%Here $\mathbf{T}^{(\ell)}(\omega)$ is the connecting matrix or the transfer matrix for the $\ell$th site.
It is clear from the above equation that the transfer matrix of the lattice $\mathbf{T}^{(l)}(\omega)$ connects the amplitude of the single particle wave-function between $\ell+n$-th site to $\ell-n$-th site which results in connecting $2n$ number of total sites. As we are dealing with clean system, the transfer matrix $\mathbf{T}^{(\ell)}(\omega)$ is independent of site $\ell$. Thus, we write the transfer matrix as $\mathbf{T}(\omega)$ instead of $\mathbf{T}^{(\ell)}(\omega)$.  
%The discrete Schr{\"o}dinger equation for finite-range hopping model $\hat{H} \psi= \omega \psi$ reads as, 
%Here $\psi_{\ell}$ is the amplitude of wave-function at $\ell$th site.
By defining, 
\begin{eqnarray}
\label{am}
a(\vert m \vert)&=&\omega/t_n, \vert m \vert =0 \nonumber \\
a(\vert m \vert) &=&t_{\vert m \vert}/t_n, \quad \vert m \vert <n ~\& ~\vert m \vert \neq 0, \nonumber \\
a(\vert m \vert)&=&1, \vert m \vert =n \quad \mathrm{and} \nonumber \\
a(\vert m \vert)&=&0,  \vert m \vert >n 
\end{eqnarray}
with $m$ denoting the $m-$th neighbor of a particular site, we can write down the general form of $2n\times 2n$ non-Hermitian transfer matrix $\mathbf{T}(\omega)$ in terms of the defined function $a(\vert m \vert)$ as~\cite{matrix_inversion},
\begin{widetext}
\begin{center}
\begin{align}
\label{general_transfer}
\mathbf{T}(\omega)=\begin{pmatrix} -a(n-1) & -a(n-2) & \ldots & - a(0) & \ldots & -a(n-2) & -a(n-1) & -1 \\
1 & 0 & \ldots & \ldots & \ldots & \ldots & 0 & 0  \\
0 & 1 & \ldots & \ldots & \ldots & \ldots & 0 & 0  \\
\vdots & \vdots & \vdots & \vdots & \vdots & \vdots & \vdots & \vdots \\
0 & 0 & \ldots & \ldots & \ldots & \ldots & 0 & 0 \\
0 & 0 & \ldots & \ldots & \ldots & \ldots & 1 & 0 
\end{pmatrix}.
\end{align}
\end{center}
\end{widetext}

We now establish the connection between the bare Green's function defined in Eq.~\ref{bare-GF} and the  transfer matrix of the lattice, introduced above in Eq.~\ref{general_transfer}. We rescale the single particle Hamiltonian $\mathbf{H}$ by $t_n$ where we recall that $t_n$ corresponds to the hopping strength of a particular site to its furthest neighbour, as allowed by the model.  
%then$th neighbour hopping, note that $n$=total number of hoppings) as, $\mathbf{H}/t_n$. 
With this rescaling, we can write down $\mathbf{g}$ as 
\begin{align}
\label{rescaledg}
\mathbf{g}(\mu)=\mathbf{\textbf{M}}(\mu)^{-1}/t_n, 
\end{align}
where 
\begin{equation}
\mathbf{\textbf{M}}(\mu)=\frac{1}{t_n} \Big[\mu \mathbb{I} - \mathbf{H}\Big]
\label{M-mat}
\end{equation} 
is a symmetric banded Toeplitz matrix which is also the case for $\mathbf{H}$ \cite{toeplitz1,toeplitz2,matrix_inversion}.
%as $\mathbf{H}$ is also a symmetric banded Toeplitz matrix with zero diagonal element. 
More explicitly, the $(i,j)$th matrix element of $\mathbf{\textbf{M}}(\mu)$ is given as
%With $ i-j =m$, the $(i,j)$th matrix element of $\textbf{M}(\mu)$ can be written as,
%\begin{align}
%\left\langle i\vert \mathbf{\textbf{M}}(\mu)\vert j\right\rangle &=a({\vert m \vert}) ;~~~\vert m \vert<n  ~~~\& ~\vert m \vert \neq 0 \\ \nonumber
%&=a(0)=\mu/t_n; ~~~~\vert m \vert = 0\\ \nonumber
%&= 1;~~~~~~\vert m \vert=n \\ \nonumber
%&=0;~~~~\vert m \vert>n
%\end{align} 

\begin{align}
\label{elementM_mu}
\left\langle i\vert \mathbf{\textbf{M}}(\mu)\vert j\right\rangle &=a({\vert m \vert})
\end{align} 
with $m=j-i$ and $a(0)=\mu/t_n$. 

As shown in Appendix.~\ref{sec:appA}, it turns out that one can write the matrix elements of $\textbf{M}(\mu)^{-1}$ given in Eq.~\ref{M-mat} in terms of the transfer matrix $\mathbf{T}(\mu)$ given in Eq.~\ref{general_transfer} as \cite{matrix_inversion}, 
\begin{widetext}
\begin{align}
\label{inverse4}
\langle i|\textbf{M}(\mu)^{-1}|j \rangle= 
\begin{cases}
   \sum\limits_{m=1}^n \left\langle n \vert \mathbf{T}(\mu)^{-i} \vert n+m \right \rangle \langle m|\textbf{M}(\mu)^{-1}|j \rangle,& 
   \text{if }~ j>i\\
    \sum\limits_{m=1}^n \left\langle n \vert \mathbf{T}(\mu)^{-i} \vert n+m \right \rangle \langle m|\textbf{M}(\mu)^{-1}|j\rangle- \left\langle n \vert \mathbf{T}(\mu)^{-(i-j+1)} \vert 1 \right\rangle,  & \text{if}~ j\leq i
\end{cases}
\end{align}
\end{widetext}
with $i, j = 1, 2, \cdots N$. Any matrix element of $\textbf{M}(\mu)^{-1}$ in  Eq.~\ref{inverse4} involves the information of $\langle m|\textbf{M}(\mu)^{-1}|j\rangle$ with $m=1,2 \ldots n$. To determine these unknown matrix elements we use the following relation (see Appendix.~\ref{sec:appA} for the details)
%we have to use some other details (see appendix A) and we can write an equation,
\begin{align}
\label{inverse5}
 \sum\limits_{m=1}^n \langle s+n \vert \mathbf{T}(\mu)^{-N} \vert n+m \rangle \, \langle m|\textbf{M}(\mu)^{-1}|j \rangle  \, \, \nonumber \\
 -\,\, \langle s+n \vert \mathbf{T}(\mu)^{-(N-j+1)} \vert 1 \rangle =0,
 \end{align}
where $s=1,2,3 \ldots n$. Therefore using Eq.~\ref{inverse5} and Eq.~\ref{inverse4} one can determine all the matrix elements of the bare Green's function $\mathbf{g}(\mu)$. Note that for the conductance calculation, we only need the component $\mathbf{g}_{1N}(\mu)$ which can be directly calculated using Eq.~\ref{inverse5}. Furthermore, it is worth noting that Eq.~\ref{inverse5} involves different powers of the transfer matrix $\mathbf{T}(\mu)$ which can be calculated by knowing the eigenspectra of the matrix. 
In section \ref{sec:TM}, we provide the relevant details on eigenvalues and eigenvectors of the transfer matrices considered here. \\

Before proceeding further, we make the following remark.
It is to be noted that without invoking the notion of transfer matrix Eq.~\ref{general_transfer}, one can directly invert the Green's function in Eq.~\ref{negf} to compute the steady-state conductance. However, such an approach does not provide a clear picture of explaining different kinds of anomalous system size scaling of conductance. An alternate promising route to capture these physics is by recasting the Green's function in Eq.~\ref{bare-GF} in terms of the underlying transfer matrix of the lattice which is inherently non-Hermitian in nature. The appearance of such transfer matrix can be understood by writing down the Schr{\"o}dinger equation for the lattice Hamiltonian as can be seen in  Eqs.~\ref{discretescro},\ref{discretescro2}, and \ref{discreteschro3}.

\section{Transfer Matrix Properties}
\label{sec:TM}
\subsection{Transfer matrix eigenvalues and its relation with lattice dispersion }
\label{subsec:eigenvalue}
In this section, we discuss the eigenvalues of the non-Hermitian transfer matrix $\mathbf{T}(\mu)$ given in Eq.~\ref{general_transfer} and its connection with the lattice dispersion relation. The characteristic equation for $\mathbf{T}(\mu)$ turns out to be,
\begin{align}
\label{eigenvalue1}
\sum\limits_{r=0}^{2n} a(\vert n-r \vert) \lambda^r =0 ,
\end{align}
where we recall that $a(\vert m \vert)$ is given in Eq.~\ref{am}. $\lambda$ denote the eigenvalues of the transfer matrix $\mathbf{T}(\mu)$. We substitute $r-n=r^{\prime}$ and rewrite Eq.~\ref{eigenvalue1} as, 
\begin{align}
\label{eigenvalue2}
\sum\limits_{r^{\prime}=-n}^{n} a(\vert r^{\prime} \vert) \lambda^{n+r^{\prime}} =0 .
\end{align}
We write the eigenvalue in the form 
\begin{equation}
\lambda=e^{i \theta}, \quad  \text{where} \,\,\, \theta \in \mathcal{C}.
\label{l-theta}
\end{equation}
The characteristic equation in Eq.~\ref{eigenvalue2} then takes the form, 
\begin{align}
\label{eigenvalue3}
e^{i n \theta} F(\theta)=0 ,
\end{align}
where we introduce, 
\begin{align}
F(\theta)=\left[ 2 \sum\limits_{r^{\prime}=1}^{n-1} a(r^{\prime}) \cos{r^{\prime} \theta} + 2 \cos{n \theta}+ a(0)\right].
\label{ftheta}
\end{align}
Since $e^{i n \theta} \neq 0$ in Eq.~\ref{eigenvalue3}, the eigenvalue spectra of the transfer matrix $\mathbf{T}(\mu)$ is obtained from the solution 
\begin{align}
\label{Ftheta}
F(\theta)=0.
\end{align}
Note that the function $F(\theta)$ defined in Eq.~\ref{Ftheta} is an even function of $\theta$. Therefore the eigenvalues of the transfer matrix (Eq.~\ref{l-theta}) always appear in the form $e^{i \theta}$ and $e^{-i \theta}$. Interestingly, if we associate the variable $\theta$ in Eq.~\ref{Ftheta} with the lattice wave-vector $k$ ($-\pi\leq k \leq \pi$), then Eq.~\ref{Ftheta} reads 
 \begin{align}
 \label{Fk}
 F(k)=0
 \end{align}
 which remarkably is the 
  dispersion relation of the finite-range hopping model in Eq.~\ref{dispersion} with $\omega(k)$ replaced by $\mu$. 
 
 For example, with nearest neighbour hopping model i.e., $n=1$, Eq.~\ref{Fk} reads,
 \begin{align}
 F(k)=2\cos k+ \frac{\mu}{t_1}=0 , \quad \mathrm{for} ~~ n=1
 \end{align}
 which gives 
 \begin{align}
 \mu=-2 t_1 \cos k, \quad ~~-\pi\leq k \leq \pi ~~\mathrm{for} ~~ n=1,
 \end{align}
 and therefore matches with the dispersion relation given in Eq.~\ref{dispersion}. The detailed discussion on finite-range hopping model with $n=2$ is given in section.~\ref{sec:result}.
\subsection{Eigenvectors of the transfer matrix $\mathbf{T}(\mu)$}
\label{subsec:eigenvector}

In this section, we provide details about the left and right eigenvectors of non-Hermitian transfer matrix $\mathbf{T}(\mu)$. Given $2n\times 2n $ transfer matrix $\mathbf{T}(\mu)$ with determinant $1$, the eigenvalues are $\lambda_k$ and $\lambda_k^{-1}$ with $k=1,2,\ldots n$.
Given an eigenvalue $\lambda_k$, the corresponding left and right eigenvectors of $\mathbf{T}(\mu)$ satisfy the following equations
\begin{eqnarray}
\langle \boldsymbol{\phi}(\lambda_k)| \mathbf{T}(\mu) &=& \lambda_k   \langle \boldsymbol{\phi}(\lambda_k)|  \nonumber \\
 \mathbf{T}(\mu) \,|\boldsymbol{\psi}(\lambda_k)\rangle &=& \lambda_k   \,|\boldsymbol{\psi}(\lambda_k)\rangle.
\end{eqnarray}
More explicitly, the left and the right eigenvectors of the transfer matrix corresponding to a given eigenvalue $\lambda_k$ can be written in a vector form as,
\begin{equation}
|{\boldsymbol{\phi}}(\lambda_k)\rangle=\begin{pmatrix}\phi_1(\lambda_k) \\
\phi_2(\lambda_k)\\
\vdots \\
\phi_{2n}(\lambda_k)
\end{pmatrix},\quad 
\mathrm{and} \quad 
|\boldsymbol{\psi}(\lambda_k)\rangle=\begin{pmatrix}\psi_1(\lambda_k) \\
\psi_2(\lambda_k)\\
\vdots \\
\psi_{2n}(\lambda_k)
\end{pmatrix}.
\end{equation}
Given the transfer matrix, $\mathbf{T}(\mu)$ in Eq.~\ref{general_transfer}, the components of the left eigenvector can be obtained as,
\begin{align}
\label{lef-eigvec}
\phi_j (\lambda_k)=\sum\limits_{r=0}^{2 n-j} a(\vert n-r \vert) \, \lambda_k^{r+j}, ~~~~j=1,2, \ldots 2n.
\end{align}
Interestingly Eq.~\ref{lef-eigvec} with $j=0$ is the characteristic polynomial for the $\mathbf{T}(\mu)$ and therefore matches with Eq.~\ref{eigenvalue1}. The similarity transformation $\mathbf{S}$ which diagonalizes the transfer matrix $\mathbf{T}(\mu)$ to its diagonal form
\begin{equation}
\label{eq:Ddiag}
\mathbf{D}=\mathrm{diag}[\lambda_1, \lambda_2, \ldots , \lambda_n \ldots \lambda_1^{-1},\lambda_2^{-1}, \ldots, \lambda_{n}^{-1}] 
\end{equation}
can be written using the left eigenvector as,
\begin{align}
\mathbf{S}=
\begin{pmatrix}
 \phi_1(\lambda_1) & \phi_2(\lambda_1) & \ldots & \phi_{2n}(\lambda_1) \\ 
  \phi_1(\lambda_2) & \phi_2(\lambda_2) & \ldots & \phi_{2n}(\lambda_2) \\
  \vdots & \vdots & \vdots & \vdots \\
   \phi_1(\lambda_n) & \phi_2(\lambda_n) & \ldots & \phi_{2n}(\lambda_n) \\ 
   \phi_1(\lambda_1^{-1}) & \phi_2(\lambda_1^{-1}) & \ldots & \phi_{2n}(\lambda_1^{-1}) \\ 
  \phi_1(\lambda_2^{-1}) & \phi_2(\lambda_2^{-1}) & \ldots & \phi_{2n}(\lambda_2^{-1}) \\
  \vdots & \vdots & \vdots & \vdots \\
   \phi_1(\lambda_n^{-1}) & \phi_2(\lambda_n^{-1}) & \ldots & \phi_{2n}(\lambda_n^{-1})
\end{pmatrix}.
\end{align}
Using the similarity transformation, we can write,
\begin{align}
\label{diagonal}
 \mathbf{S}\,\mathbf{T}(\mu)\,\mathbf{S}^{-1}=\mathbf{D}. 
\end{align}
Here, $\mathbf{S}^{-1}$ contains all the components of right eigenvector as,
\begin{align}
\mathbf{S}^{-1}=\begin{pmatrix}
 \psi_1(\lambda_1)  & \ldots &  \!\!\!\!\psi_1(\lambda_n)& \!\psi_1(\lambda_1^{-1})  & \ldots & \!\!\!\!\psi_1(\lambda_n^{-1})\\ 
 \psi_2(\lambda_1)  & \ldots & \!\!\!\!\psi_2(\lambda_n)& \!\psi_2(\lambda_1^{-1})  & \ldots & \!\!\!\!\psi_2(\lambda_n^{-1})\\ 
 \vdots & \vdots & \vdots & \vdots & \vdots & \vdots \\
 \!\psi_{2n}(\lambda_1)  & \ldots & \!\!\!\!\psi_{2n}(\lambda_n)& \psi_{2n}(\lambda_1^{-1})  & \ldots & \!\!\!\!\psi_{2n}(\lambda_n^{-1}) 
\end{pmatrix}.
\end{align}

%The proof of this relation is given in Appendix A.
\subsection{Matrix elements of the exponents of transfer matrix $\mathbf{T}(\mu)$}
\label{subsec:exponents}

In this section, we compute the exponents of the transfer matrix that are required to obtain the conductance. If the transfer matrix $\mathbf{T}(\mu)$ is diagonalizable, then using Eq.~\ref{diagonal} we can write the $m-$th exponent of $\mathbf{T}(\mu)$ as,
\begin{equation}
\mathbf{T}^m(\mu)= \mathbf{S}^{-1} \, \mathbf{D}^m \, \mathbf{S}. 
\end{equation}
Thus, the  matrix elements of the $m-$th exponent $\mathbf{T}(\mu)$ is given by,
\begin{eqnarray}
\label{matrixexponent}
&&\left\langle s \vert \mathbf{T}^{m}(\mu) \vert j \right \rangle =\sum\limits_{k=1}^ {2n} \psi_s(\lambda_k)\, \phi_j (\lambda_k) \, \lambda_k^m   \nonumber \\
&&=\sum\limits_{k=1}^n \Big[ \psi_s(\lambda_k)\,\phi_j (\lambda_k)\, \lambda_k^m+ \psi_s(\lambda_k^{-1})\, \phi_j (\lambda_k^{-1})\,\lambda_k^{-m}\Big]. \nonumber \\
\end{eqnarray}
In cases when the transfer matrix $\mathbf{T}(\mu)$ is no longer diagonalizable, one can bring it to a Jordan-normal form. One such situation arises when at least two eigenvalues are the same. Generally, this does not necessarily imply coalescing of two eigenvectors. However, in the case of transfer matrix $\mathbf{T}(\mu)$, the analytical mathematical structure  (Eq.~\ref{general_transfer}) facilitates one to recast the eigenvectors in the form of Eq.~\ref{lef-eigvec}. It is interesting to note coalescing of eigenvalues in Eq.~\ref{lef-eigvec} also necessarily implies  coalescing of eigenvectors.  If $\mathbf{R}$ is any similarity transformation that converts  $\mathbf{T}(\mu)$  to a $2n\times 2n$ Jordan-normal form $J$ then 
\begin{equation}
    \mathbf{R} \, \mathbf{T}(\mu) \, \mathbf{R}^{-1}=\mathbf{J}.
    \label{Jordan}
\end{equation} 
%where $\mathbf{J}$ is a $2n\times 2n$ matrix with Jordan-normal form. 
As a result, $\mathbf{T}(\mu)=\mathbf{R}^{-1}\, \mathbf{J} \, \mathbf{R}$. Thus, in this case, we can calculate the exponents as, 
\begin{equation}
\mathbf{T}^m(\mu)= \mathbf{R}^{-1} \, \mathbf{J}^m \, \mathbf{R}. 
\end{equation}
Up to now, all the descriptions are very general for the finite-range hopping model. In the result section, we will describe the specific examples in detail and look at the connection between non-Hermitian properties of transfer matrix $\mathbf{T}(\mu)$ and its relation with open-system conductance.

\section{Results}
\label{sec:result}
In this section, we present our results for the finite-range lattice model with a range of hopping $n=2$ and unravel the important novel role played by the eigenspectra of the transfer matrix in comparison to the nearest neighbour case i.e., $n=1$. Before proceeding further, we would like to list certain important and pertinent questions: 

\begin{figure}
\includegraphics[width=\columnwidth]{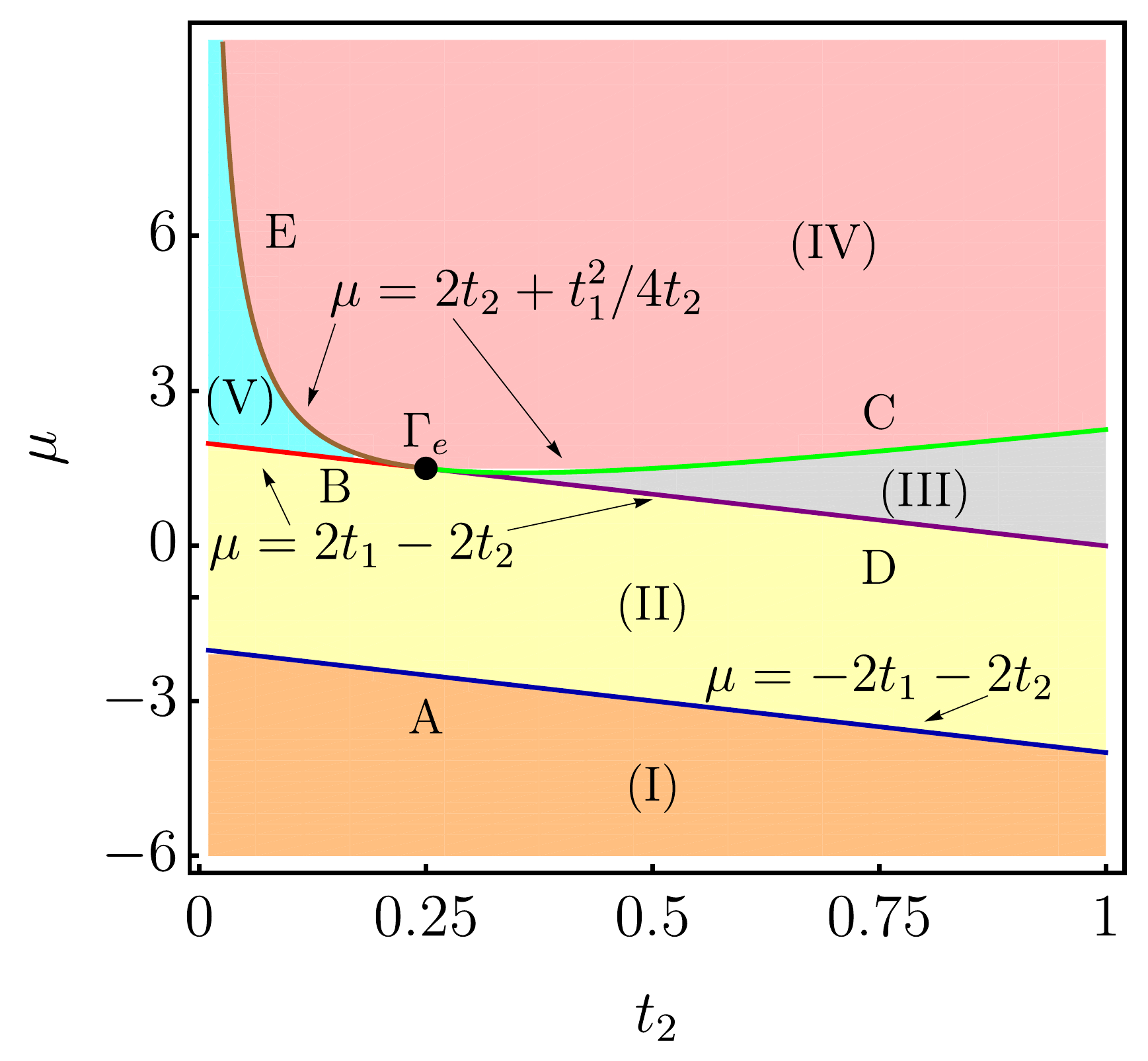} 
\caption{The figure represents a schematic phase diagram in the $\mu-t_2$ plane. We fix $t_1=1$ without loss of generality. The phase diagram is constructed using the eigenvalues of transfer matrix $\mathbf{T}(\mu)$ for $n=2$. We identify five different regimes denoted by (I) to (V) (orange, yellow, gray, pink, and cyan) where each regime is associated by the nature of complex eigenvalues. This is further elaborated in Table.~\ref{table1}. The lines between the regimes denoted by A-E (blue, red, green, purple, and brown) represents the exceptional lines. In other words, at any given point on any of these lines at least two of the four eigenvalues and eigenvectors coalesce. Remarkably, a special point emerges in the phase plane as a result of intersection between two exceptional lines and occurs at $t_2=t_1/4$. This point is denoted by $\Gamma_e$ (black filled circle) and all four eigenvalues and eigenvectors evaluated at this point coalesce indicating a fourth order exceptional point. The blue line (A) corresponds to the lower band edge as given in Eq.~\ref{lower band edge}. Similarly the red line (B), the fourth order exceptional point  $\Gamma_e$ and the green line (C) form the upper band edge given in Eq.~\ref{resultn=2}. The other two lines i.e. purple line (D) and brown line (E) albeit being  exceptional point interestingly do not correspond to the band edges. The NESS conductance inside the five regimes [(I)-(V)], on the five lines (A to E) and the higher order exceptional point ($\Gamma_e$) has been discussed in Table.~\ref{table1} and ~\ref{table2}. } 
\label{schematic}
\end{figure}

%Recall that, the conductance 

%Note that, the case $n=1$ i.e., the nearest neighbour hopping model has recently been studied 

%Conductance and its relation with non-Hermitian transfer matrix properties of finite-range hopping model with $n=1$ (nearest neighbour hopping model) have been studied recently. Exceptional point of transfer matrix is a great tool to  describe the universal subdiffusive transport ($1/N^2$ scaling) behaviour at the band edges for this case. 

%It has been also reported that though the transfer matrix picture does not work for the long-range hopping model (here number of hopping scales with system size), at the band edges of the long-range hopping model conductance still shows the universal subdiffusive decay ($1/N^2$). Now, for the finite-range hopping model with $n>1$ and $n \ll N$, the transfer matrix picture still works. 

\begin{enumerate}
    \item Do the band edges for the finite-range hopping model correspond to the exceptional points of the transfer matrix? If yes, what is the consequence in terms of NESS transport? 
    
    \item As discussed in Sec.~\ref{sec:LH_DR} for finite range hopping model with $n>1$, the upper band edge may or may not correspond to $k=\pi$. What is the corresponding signature, if any, in NESS transport?

    \item With increasing the hopping range, the dimension of the transfer matrix also increases. Do these transfer matrices support exceptional hyper-surfaces with higher-order exceptional points and if yes, are their consequences in NESS transport?
    
\end{enumerate}
To answer these questions, we now discuss a concrete example of finite-range hopping model with $n=2$ (next nearest neighbour hopping model), without loss of generality and comment on the case of general $n$.

\subsection{An example of finite-range hopping model with $n=2$}
\label{n-2-sec}
{\it Dispersion and band edges:}
The dispersion relation for the finite range hopping model is given in Eq.~\ref{dispersion}. For $n=2$, with nearest neighbour hopping $t_1$ and next nearest neighbour hopping $t_2$, we get, 
\begin{equation}
\omega(k)=-2 t_1 \cos k- 2 t_2 \cos 2k.
\end{equation}
Recall that, the lower band edge is always at $k=0$ and the corresponding energy is given by 
\begin{equation}
    \omega(k\!=\!0)=-2 t_1 - 2t_2, \quad \text{lower band edge}
    \label{lower-2}
\end{equation} 
In contrast, whether $k=\pi$ is an upper band edge or not, depends on a condition between the two hoppings, as discussed for the general case in Sec.~\ref{sec:LH_DR}. The energy of the upper band edge for three different scenarios mentioned in Eq.~\ref{condition1}, Eq.~\ref{condition2}, and Eq.~\ref{condition3} is given by,

\begin{align}
\label{resultn=2}
\omega(k)= 
\begin{cases}
   2 t_1-2 t_2,& 
   \text{if }~ t_2 < t_1/4~~\textrm{with}~ k=\pi\\
    2 t_1-2 t_2,  & \text{if}~ t_2 =t_1/4~~\textrm{with}~ k=\pi \\
    \frac{t_1^2}{4 t_2}+ 2t_2, & \text{if}~ t_2 >t_1/4~~\textrm{with}~ k=\cos^{-1}[-\frac{t_1}{4t_2}].
\end{cases}
\end{align}

Next, we discuss in detail the nature of the eigenvalues of the transfer matrix $\mathbf{T}(\mu)$.

%For $t_2\leq t_1/4$, the value of the upper band edge is $\omega(k)=2 t_1-2 t_2$ and this is for $k=\pi$. For $t_2> t_1/4$, the upper band edge will correspond to $k=\cos^{-1}[-\frac{t_1}{4t_2}]$ and the value of upper band edge is $\omega(k)=\frac{t_1^2}{4 t_2}+ 2t_2$. Thus, the three different scenarios for $n=2$ reads as (a) $t_2<t_1/4$ (b) $t_2>t_1/4$ and (c) $t_2=t_1/4$. 
\begin{table}
\begin{tabular}{| c | c | c | c |}
\hline 
 Regimes  & $\mu$ &  Transfer matrix                    &      Conductance        \\
  &   & eigenvalues & $\mathcal{G}(\mu)$  \\
\hline
Below           &    $\mu<-2t_1-2t_2$  &  $ \lambda_1=-e^{\kappa_1},$    &    $e^{-N/\xi}$                       \\
lower & & $\lambda_2=e^{\kappa_2},$& No transport \\
 band edge&  &  $ \lambda_1^{-1}=-e^{-\kappa_1}, $   &  \\
(I) & & $\lambda_2^{-1}=e^{-\kappa_2}$& \\
\hline
Within  & $-2t_1-2t_2<\mu$ & $\lambda_1=-e^{\kappa_1}$, & $N^0$ \\
band edge & $< 2 t_1 -2 t_2$&  $\lambda_2=e^{i\kappa_2}$, & Ballistic\\
 (II) & & $\lambda_1^{-1}=-e^{-\kappa_1},$& \\
& &$\lambda_2^{-1}=e^{-i\kappa_2}$ & \\
\hline
 Within  &$2 t_1 -2 t_2<\mu$ & $ \lambda_1=e^{i\kappa_1},$  & $N^0$ \\
 band edges   &$<2 t_2 + \frac{t_1^2}{4 t_2};$ &$\lambda_2=e^{i\kappa_2},$   & Ballistic\\
 (III) &(valid only  &$ \lambda_1^{-1}=e^{-i\kappa_1},$ & \\
  & when $t_2>t_1/4$) &$\lambda_2^{-1}=e^{-i\kappa_2}$  & \\

\hline
Above & $\mu>2 t_2 + \frac{t_1^2}{4 t_2}$& $ \lambda_1=-e^{\kappa_1+ i \kappa_2}$,& $e^{-N/\xi}$ \\
 upper& & $\lambda_2=-e^{\kappa_1-i \kappa_2}$, & No transport \\
 band edge& & $ \lambda_1^{-1}=-e^{-\kappa_1- i \kappa_2},$ &  \\
(IV) & & $ \lambda_2^{-1}=-e^{-\kappa_1+ i \kappa_2}$ & \\
 \hline
Above  &$2 t_1 -2 t_2<\mu$ & $ \lambda_1=-e^{\kappa_1},$  & $e^{-N/\xi}$ \\
 upper   &$<2 t_2 + \frac{t_1^2}{4 t_2}$ &$\lambda_2=-e^{\kappa_2},$   & No transport\\
 band edge &(valid only &$ \lambda_1^{-1}=-e^{-\kappa_1},$ & \\
 (V) & when $t_2<t_1/4$)&$\lambda_2^{-1}=-e^{-\kappa_2}$  & \\
\hline   
 \end{tabular}
\caption{The table represents details of various properties of regimes [(I)-(V)] shown in Fig.~\ref{schematic}. The second column gives the allowed values of chemical potential $\mu$ in these regimes. The third column gives the eigenvalues of the transfer matrix $\mathbf{T}(\mu)$ evaluated at the chemical potential $\mu$. One can notice the change in the nature of the eigenvalues when moving from one regime to another regime. The last column gives the corresponding system size scaling of NESS conductance $\mathcal{G}(\mu)$. It is important to emphasize that no exceptional lines or points appear inside these regimes and hence the transfer matrix is always diagonalizable. The scenario in which exceptional lines or points appear is discussed in Table.~\ref{table2}. }
\label{table1}
\end{table}

\begin{table}
\begin{tabular}{| c | c | c | c |}
\hline 
 Exceptional  & $\mu$ &  Transfer matrix                    &      Conductance        \\
 lines/points &   & eigenvalues & $\mathcal{G}(\mu)$  \\
\hline
Lower        & $\mu=-2t_1-2t_2$      &    $ \lambda_1=1,$      &    $1/N^2$                        \\
band edge      & & $\lambda_2=-e^{\kappa_1},$  &  subdiffusive    \\
A (EL) & & $\lambda_1^{-1}=1,$& \\
&  & $ \lambda_2^{-1}=-e^{-\kappa_1}$,  & \\
\hline
Upper        & $\mu=2t_1-2t_2$      &    $ \lambda_1=-1,$      &    $1/N^2$                        \\
band edge      & & $\lambda_2=-e^{\kappa_1},$  &  subdiffusive    \\
B (EL)&(valid only & $\lambda_1^{-1}=-1,$& \\
& when $t_2<t_1/4$) & $ \lambda_2^{-1}=-e^{-\kappa_1}$  & \\
\hline
 Upper& $\mu=2 t_1 -2 t_2$&$ \lambda_1=-1,$  & $1/N^2$  \\
 band edge & &  $ \lambda_2=-1,$  &  subdiffusive\\
 $\Gamma_e$ (fourth & (valid only& $ \lambda_1^{-1}=-1,$ & \\
 order EP) & when $t_2=t_1/4$) & $\lambda_2^{-1}=-1$  &  \\
 \hline
  Upper&$\mu=2 t_2 + \frac{t_1^2}{4 t_2}$ &$ \lambda_1=e^{i\kappa_1},$   & $1/N^2$ \\
 band edge& & $\lambda_2=e^{-i\kappa_1},$ &subdiffusive \\
 C (EL)&(valid only &$ \lambda_1^{-1}=e^{-i\kappa_1},$ & envelope with \\
 &when $t_2>t_1/4$) & $\lambda_2^{-1}=e^{i\kappa_1}$ & oscillations \\
 \hline
 Within& $\mu=2 t_1 -2 t_2$&$ \lambda_1=-1,$  & $N^0$  \\
 band edge& &  $ \lambda_2=e^{i\kappa_1},$  &  Ballistic\\
D (EL)&(valid only & $ \lambda_1^{-1}=-1,$& \\
 & when $t_2>t_1/4$)& $\lambda_2^{-1}=e^{-i\kappa_1}$  &  \\
 \hline
  Above & $\mu=2 t_2 + \frac{t_1^2}{4t_2}$&$ \lambda_1=-e^{\kappa_1},$  & $e^{-N/\xi}$  \\
 upper& &  $ \lambda_2=-e^{-\kappa_1},$  &  No transport\\
band edge& (valid only& $ \lambda_1^{-1}=-e^{-\kappa_1},$& \\
 E (EL)& when $t_2<t_1/4$) & $\lambda_2^{-1}=-e^{\kappa_1}$  &  \\
 \hline
\end{tabular}
\caption{This table represents the scenario when transfer matrix is not diagonalizable and this occurs at exceptional lines (EL) or points (EP). The first column gives the different exceptional lines (A-E) and a higher order exceptional point ($\Gamma_e$) shown in Fig.~\ref{schematic}. The second column gives the allowed values of chemical
 potential $\mu$ in these lines and point. The third column gives the eigenvalues of the transfer matrix $\mathbf{T}(\mu)$ evaluated at the chemical potential $\mu$. The last column gives the corresponding system size scaling of NESS conductance $\mathcal{G}(\mu)$.}
\label{table2}
\end{table}

\begin{figure*}
\includegraphics[width=2\columnwidth]{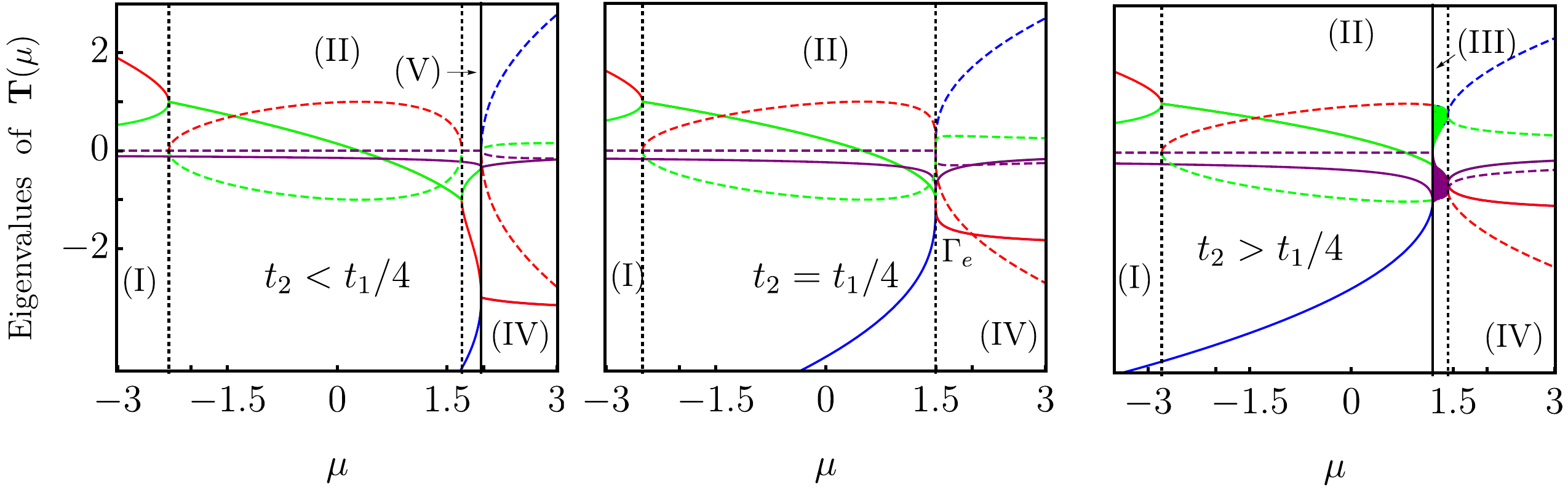} 
\caption{Plots for real (solid line) and imaginary parts (dashed line) of the eigenvalues of $4 \times 4$ transfer matrix $\mathbf{T}(\mu)$ for $n=2$ as a function of $\mu$. We set the value $t_1=1$.  [Left panel] Here we consider the case $t_2< t_1/4$, with $t_2=0.15$. This is an example of making a vertical cut in Fig.~ \ref{schematic} in the zone where $t_2<0.25$. Note that such a cut in Fig.~ \ref{schematic}  passes through four regimes (I), (II), (V) and (IV) which are also marked here. Additionally, this cut encompasses three exceptional points out of which two correspond to the lower and the upper band edges. Recall that these lower and upper band edges are denoted by $A$ and $B$, respectively in Fig.~ \ref{schematic}. The other exceptional point which does not correspond to any band edges is denoted by $E$. In this figure, we represent the exceptional points corresponding to band edges by black dotted vertical lines and the exceptional points that do not correspond to the band edges by black solid vertical lines. The nature of the eigenvalues is consistent with that summarized in Table.\ref{table1}. In regime (I), the eigenvalues (real part) represented by red and green solid lines are inverse of each other. Likewise, there is an inverse corresponding to the purple line which, for the sake of clarity, is not represented here as it falls way outside the presented $y$-axis range. In the same manner, even in other regimes/cases when data values fall outside the presented $y$-axis range, we do not present it here. [Middle panel] Here we consider the case $t_2=t_1/4=0.25$ which spans three regimes  (I), (II), and (IV) and two exceptional points corresponding to two band edges according to  Fig.~\ref{schematic}. Note that, the exceptional point corresponding to the upper band edge is denoted by $\Gamma_{e}$ which is a fourth-order exceptional point. As can be seen in the figure, at the point $\Gamma_e$, all four eigenvalues become real and coalesce at the value $-1$. [Right panel] (c) Here we consider the case $t_2>t_1/4$ with $t_2=0.4$. Similar to the top panel, a corresponding appropriate vertical cut passes through regimes (I), (II), (III), and (IV). It is worth noting that regime (III) is characterized by a scenario where all four eigenvalues are complex with absolute value $1$. As a consequence, the eigenvalues when plotted as a function of $\mu$ in this regime appear dense.}
%The dotted vertical lines in black  correspond to exceptional points of the transfer matrix that occur at the upper and the lower band edges. The solid vertical line in black corresponds to the transfer matrix exceptional point that does not occur at the band edges.} 
%Here, the two vertical dotted lines are at the band edges and also have exceptional points.} 
\label{eigenvalues1} 
\end{figure*}

{\it Detailed description of nature of eigenvalues of transfer matrix $\mathbf{T}(\mu)$:}
The transfer matrix $\mathbf{T}(\mu)$ for $n=2$ is a $4\times 4$ matrix and is given by (using Eq.~\ref{general_transfer}),
\begin{align}
\mathbf{T}(\mu)=\begin{pmatrix} -\frac{t_1}{t_2} & -\frac{\mu}{t_2} &- \frac{t_1}{t_2}& -1 \\
1& 0 & 0& 0 \\
0 & 1 & 0 & 0\\
0& 0 & 1 & 0\end{pmatrix}.
\label{2T}
\end{align}
We now unravel the properties of this transfer matrix. In Fig.~\ref{schematic}, we first construct a phase diagram in the $\mu-t_2$ plane, by setting $t_1=1$ without loss of generality. The phase diagram is constructed by using the different nature of the eigenvalues of $\mathbf{T}(\mu)$. The eigenvalues of $\mathbf{T}(\mu)$ are obtained analytically following Eq.~\ref{Ftheta} (see Appendix.~\ref{sec:app2} for the details).  It turns out that the lower band edge, as given in Eq.~\ref{lower-2}, is an exceptional line in the $\mu-t_2$ plane. This implies that at any given point on this line at
least two of the four eigenvalues and eigenvectors of $\mathbf{T}(\mu)$ coalesce. This line is denoted by the symbol A (blue solid line). Similarly, following Eq.~\ref{resultn=2}, the upper band edge also turns out to yield  exceptional lines denoted by B (purple solid line) and C (green solid line) and a higher order exceptional point denoted by $\Gamma_e$ (black filled circle).  The other two lines in the phase plane i.e. D (purple solid line) and E (brown solid line) do not correspond to band edges although remarkably still remain as exceptional lines. These exceptional lines A-E yield five different regimes, denoted by (I) to (V) [orange, yellow, gray,
pink, and cyan]. This Fig.~\ref{schematic} sets the stage for a more quantitative summary of our main findings which is gathered in Table \ref{table1} and Table \ref{table2}. The different regimes and associated properties of $\mathbf{T}(\mu)$ are described in Table \ref{table1}.  In Table \ref{table2} we present the findings for different exceptional lines and the higher order exceptional point.  In addition  to the properties of $\mathbf{T}(\mu)$, we further summarize in the last columns of Table \ref{table1} and Table \ref{table2} the system size scaling of NESS conductance which we will be discussed in depth later.  The third column of Table \ref{table1} and Table \ref{table2} can be nicely visualized via appropriate vertical cuts in Fig.~\ref{schematic}. This is presented in detail in Fig.~\ref{eigenvalues1}.

%---------------------Table 1--------------------------------

In what follows, we first present the direct numerical results for the NESS conductance ${\mathcal G}(\mu)$ following Eq.~\ref{conductance} and Eq.~\ref{negf} and its system size scaling at different regimes (I) - (V) and at the exceptional lines/point. We further provide an in-depth analysis of these scalings in terms of the transfer matrix eigenspectra.

\begin{figure*}
\includegraphics[width=2\columnwidth]{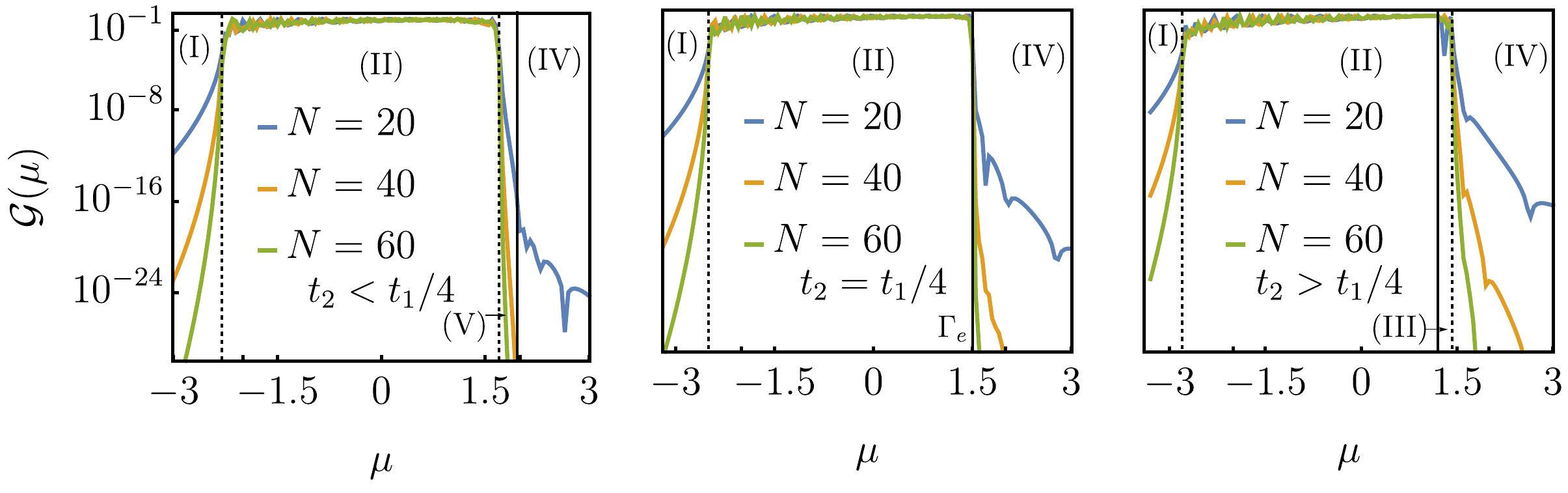} 
\caption{Plot for conductance $\mathcal{G}(\mu)$ as a function of $\mu$ for three different and appropriate vertical cuts ($t_2 < t_1/4$, $t_2 = t_1/4$, $t_2 > t_1/4$) in the $\mu-t_2$ phase plane in Fig.~\ref{schematic} for different system sizes $N$. We set the value $t_1$ and $t_2$ exactly the same as in Fig.~\ref{eigenvalues1}. In all the figures, we represent the exceptional points corresponding to band edges by black dotted vertical lines and the exceptional
points that do not correspond to the band edges by black solid vertical lines. In all cases, we see non-analytic changes in $\mathcal{G}(\mu)$ at the two band edges and this is discussed in more detail in Fig.~\ref{cond3}. [Left panel] Here we consider the case $t_2 < t_1/4$ with $t_1=1$ and $t_2=0.15$. The behavior of conductance in four different regimes (I), (II), (V), and (IV) are shown. In regimes (I), (V), and (IV) which correspond to outside the band edges, the conductance decays exponentially with system size $N$. In regime (II) which corresponds to within the band edges, ballistic behavior (system size independence) is observed. [Middle panel] Here we consider the case $t_2 = t_1/4$ with $t_1=1$ and $t_2=0.25$. Regime (II) shows ballistic transport and regimes (I) and (IV) show exponentially suppressed transport.   [Right panel] Here we consider the case $t_2 > t_1/4$ with $t_1=1$ and $t_2=0.4$. Regimes (II) and (III) show ballistic transport and regimes (I) and (IV) show exponentially suppressed transport.}
\label{cond1} 
\end{figure*}

\begin{figure*}
\includegraphics[width=2\columnwidth]{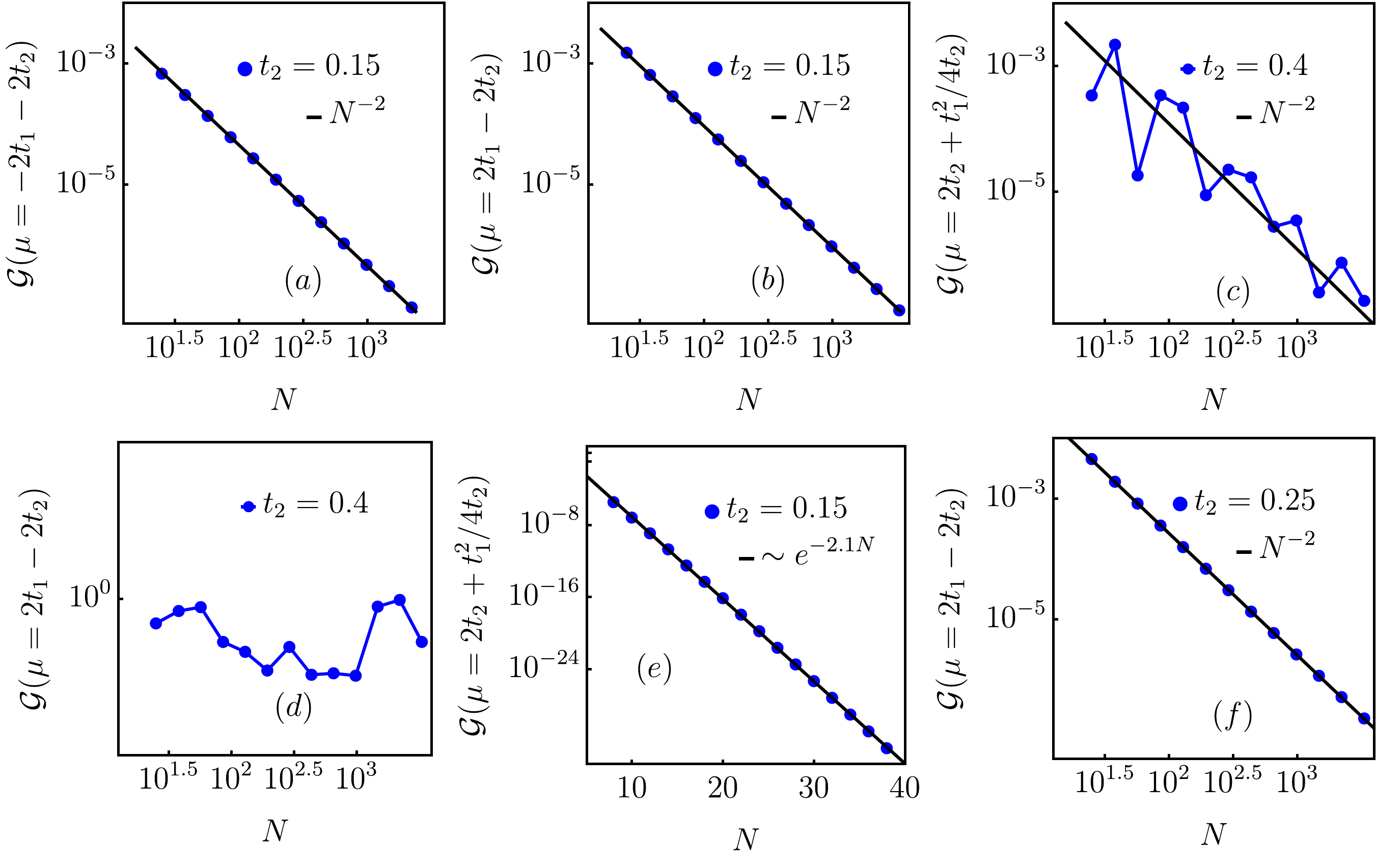} 
\caption{Plot for system size scaling of NESS conductance  $\mathcal{G}(\mu)$ at exceptional lines A-E and point $\Gamma_e$ as elucidated in Fig.~\ref{schematic}. (a) We choose a point that lies on line $A$ (lower band edge) in Fig.~\ref{schematic} and observe subdiffusive scaling  $\mathcal{G}(\mu) \sim 1/N^2$. (b) We choose a point that lies on line $B$ (upper band edge with $t_2<t_1/4$) and once again observe subdiffusive scaling with the same exponent. (c) We choose a point that lies on line $C$ (upper band edge with $t_2>t_1/4$) which exhibits interesting oscillating behviour with an overall $1/N^2$ envelope. This therefore can also be regarded as an anomalous behaviour that is subdiffusive in nature. (d) We choose a point that lies on line $D$ (within the band edges with $t_2 >t_1/4$) and observe system size independent scaling (ballistic transport) despite being an exceptional point. (e) We choose a point that lies on line $E$ (outside the upper band edges with $t_2 < t_1/4$) and observe exponentially decaying scaling of conductance with system size despite being an exceptional point. (f) We choose the $\Gamma_e$ point which occurs at $t_2=t_1/4$ and corresponds to upper band edge. We observe subdiffusive transport with the same exponent despite being a higher-order exceptional point. These six plots reveal that the cause of the anomalous transport is rooted in the existence of exceptional points at the band edges.} 
%at the upper band-edge for three different scenarios given in E%q.~\ref{resultn=2}. Here (a) $t_2<t_1/4$, (b) $t_2=t_1/4$, and (c) $t_2>t_1/4$ with parameter values chosen same as Fig.~\ref{cond1}. In all cases, conductance scales subdiffusively with scaling $N^{-2}$. In (c), an interesting oscillatory feature is observed with an overall envelope that scales as $N^{-2}$, which is rooted in the scenario where the upper band edge is not located at $k=\pi$.}
%we can see a subdiffusive decay of conductance as $1/N^2$ with oscillations. Fig. (c), corresponds to the scaling behavior for $t_2=t_1/4$. Here, in plot (a) and (c), we can see the  subdiffusive decay of conductance as $1/N^2$ without any oscillations. } 
\label{cond3} 
\end{figure*}

{\it{NESS conductance and its scaling with system size:}}
In Fig.~\ref{cond1}, we show the results for conductance $\mathcal{G}(\mu)$ as a function of $\mu$ for different system sizes $N$. Different relative values of $t_1$ and $t_2$ are chosen. These correspond to appropriate vertical cuts ($t_2 < t_1/4$, $t_2 = t_1/4$, $t_2 > t_1/4$) in the $\mu-t_2$ phase plane in Fig.~\ref{schematic}. In all three cases, $\mathcal{G}(\mu)$ displays non-analytic changes at both upper and lower band edges.  It is evident that within the band edges  i.e., regime (II) in the left and middle panels and regime (II) and (III) in the right panel, $\mathcal{G}(\mu)$ is independent of system size and therefore implies ballistic transport. It is worth noting that the black solid vertical line that separates regimes (II) and (III) in the right panel of Fig.~\ref{cond1}, represents an exceptional point on the line (D) of Fig.~\ref{schematic}. Despite this point being exceptional, ballistic transport behavior is observed and is shown in Fig.~\ref{cond3}(d). Outside of both the lower and upper band edges, i.e., regime (I), (V), and (IV) in the left panel, regimes (I) and (IV) in the middle and last panels of Fig.~\ref{cond1}, $\mathcal{G}(\mu)$ decays exponentially with system size. Once again, the black solid vertical line that separates regimes (V) and (IV) in the left panel of Fig.~\ref{cond1}, despite being exceptional, displays exponentially decaying transport and is shown in Fig.~\ref{cond3}(e).

In striking contrast, at both the band edges which correspond to exceptional lines A, B, and C or point $\Gamma_e$ as elucidated in Fig.~\ref{schematic}, $\mathcal{G}(\mu)$ shows interesting anomalous transport with $1/N^2$ scaling.  This is clearly demonstrated in Fig.~\ref{cond3} (a), (b), (c), and (f). Moreover, in Fig.~\ref{cond3}(c) in addition to overall, $1/N^2$ envelope, there are interesting oscillations whose cause is rooted in the fact that exceptional point albeit occurring at the band edge does not correspond to $k=\pi$ (see Eq.~\ref{resultn=2}). Recall that albeit Fig.~\ref{cond3}(f) is associated with a fourth-order exceptional point $\Gamma_e$, nonetheless, the robustness of $1/N^2$ scaling of NESS conductance is observed, indicating remarkable universality in anomalous transport. Our findings provide strong evidence that the cause of anomalous transport is rooted not only in the existence of exceptional points but also in the fact that they have to be associated with the band edges.  In what follows, we bolster our findings by using suitable analytical arguments based on transfer matrices. 
%explain all these scaling behaviours in terms of the non-Hermitian transfer matrix corresponding to $n=2$ and its exceptional points.

{\it{Analytical approach to NESS conductance scaling in terms of non-Hermitian transfer matrix:}} As mentioned earlier, the system size scaling of NESS conductance $\mathcal{G}(\mu)$ is entirely governed by the bare Green's function of the system i.e., $\vert \mathbf{g}_{1N}(\mu)\vert ^2$ with $\mathbf{g}(\mu)$ being defined in Eq.~\ref{bare-GF}. Using Eq.~\ref{inverse5} the $\mathbf{g}_{1N}(\mu)$ component can be obtained via the transfer matrix.  Recall that $\mathbf{g}(\mu)$ is related to $\textbf{M}(\mu)^{-1}$ via Eq.~\ref{rescaledg}. According to Eq.~\ref{inverse5}, $\textbf{M}(\mu)^{-1}$ obeys the following relations for $n=2$,
\begin{widetext}
\begin{align}
\label{eq:g1N}
 \left\langle 3 \vert \mathbf{T}(\mu)^{-N} \vert 3 \right \rangle \left\langle 1 \vert\textbf{M}(\mu)^{-1}\vert j \right\rangle+ \left\langle 3 \vert \mathbf{T}(\mu)^{-N} \vert 4 \right \rangle \left\langle 2\vert\textbf{M}(\mu)^{-1}\vert j \right\rangle 
 - \left\langle 3 \vert \mathbf{T}(\mu)^{-(N-j+1)} \vert 1 \right\rangle =0.  \\ \nonumber 
  \left\langle 4 \vert \mathbf{T}(\mu)^{-N} \vert 3 \right \rangle \left\langle 1 \vert\textbf{M}(\mu)^{-1}\vert j \right\rangle+ \left\langle 4 \vert \mathbf{T}(\mu)^{-N} \vert 4 \right \rangle \left\langle 2 \vert\textbf{M}(\mu)^{-1}\vert j \right\rangle- \left\langle 4 \vert \mathbf{T}(\mu)^{-(N-j+1)} \vert 1 \right\rangle =0.
\end{align}
Eq.~\ref{eq:g1N} can be recast to a matrix form, 
\begin{align}
\label{eq:g-mat}
  \begin{pmatrix}
  \left\langle 3 \vert \mathbf{T}(\mu)^{-N} \vert 3 \right \rangle & \left\langle 3 \vert \mathbf{T}(\mu)^{-N} \vert 4 \right \rangle \\
  \left\langle 4 \vert \mathbf{T}(\mu)^{-N} \vert 3 \right \rangle & \left\langle 4 \vert \mathbf{T}(\mu)^{-N} \vert 4 \right \rangle \end{pmatrix} \begin{pmatrix}
  \left\langle 1 \vert\textbf{M}(\mu)^{-1}\vert j \right\rangle \\
  \left\langle 2 \vert\textbf{M}(\mu)^{-1}\vert j \right\rangle
  \end{pmatrix} = \begin{pmatrix}
  \left\langle 3 \vert \mathbf{T}(\mu)^{-(N-j+1)} \vert 1 \right\rangle \\
  \left\langle 4 \vert \mathbf{T}(\mu)^{-(N-j+1)} \vert 1 \right\rangle
  \end{pmatrix}.
\end{align}
As for the NESS conductance calculation, we require $\mathbf{g}_{1N}(\mu)$ component which in turn requires us to evaluate $\langle 1| \textbf{M}(\mu)^{-1} | N \rangle$. We therefore set $j=N$ and obtain from Eq.~\ref{eq:g-mat}
\begin{align}
\label{g1na}
 \begin{pmatrix}
  \left\langle 1 \vert\textbf{M}(\mu)^{-1}\vert N \right\rangle \\
  \left\langle 2 \vert\textbf{M}(\mu)^{-1}\vert N \right\rangle
  \end{pmatrix}=  \begin{pmatrix}
  \left\langle 3 \vert \mathbf{T}(\mu)^{-N} \vert 3 \right \rangle & \left\langle 3 \vert \mathbf{T}(\mu)^{-N} \vert 4 \right \rangle \\
  \left\langle 4 \vert \mathbf{T}(\mu)^{-N} \vert 3 \right \rangle & \left\langle 4 \vert \mathbf{T}(\mu)^{-N} \vert 4 \right \rangle \end{pmatrix}^{-1}  \begin{pmatrix}
  \left\langle 3 \vert \mathbf{T}(\mu)^{-1} \vert 1 \right\rangle \\
  \left\langle 4 \vert \mathbf{T}(\mu)^{-1} \vert 1 \right\rangle
  \end{pmatrix}.
\end{align}
\end{widetext}
Using Eq.~\ref{g1na}, one can easily evaluate $\mathbf{g}_{1N}(\mu)=\left\langle 1 \vert\textbf{M}(\mu)^{-1}\vert N \right\rangle/t_2$ with $\mathbf{T}(\mu)$ as given in Eq.~\ref{2T}.  By performing the inverse of  $\mathbf{T}(\mu)$ in Eq.~\ref{2T}, it is easy to check that $\left\langle 3 \vert \mathbf{T}^{-1} (\mu)\vert 1 \right \rangle=0$ and $\left\langle 4 \vert \mathbf{T}^{-1}(\mu)\vert 1 \right \rangle=-1$. To this end, we obtain a simplified expression for $\mathbf{g}_{1N}(\mu)$ as 
\begin{align}
\label{eqcond1}
\mathbf{g}_{1N}=\frac{1}{t_2}\frac{  \left\langle 3 \vert \mathbf{T}^{-N} \vert 4 \right \rangle  }{\left\langle 4 \vert \mathbf{T}^{-N} \vert 4 \right \rangle \left\langle 3 \vert \mathbf{T}^{-N} \vert 3 \right \rangle - \left\langle 3 \vert \mathbf{T}^{-N} \vert 4 \right \rangle \left\langle 4 \vert \mathbf{T}^{-N} \vert 3 \right \rangle}.
\end{align}
For the sake of brevity we omit the argument $\mu$ from both $\mathbf{T}$ and $\mathbf{g}_{1N}$ in Eq.~\ref{eqcond1}. 

Eq.~\ref{eqcond1} is one of the central equations of this  work. Thus, to calculate $\vert \mathbf{g}_{1N}\vert ^2$, the main task is to calculate $\mathbf{T}^{-N}$ using its eigenspectra. Recall that, in Table.~\ref{table1} and Table~\ref{table2}, we summarize the nature of the eigenvalues of the transfer matrix according to the different regimes of Fig.~\ref{schematic}. From the nature of these eigenvalues, one can extract the system size dependence of NESS conductance $\mathcal{G}(\mu)$ using Eq.~\ref{eqcond1}, as we discuss below.

Let us now consider a situation when the transfer matrix $\mathbf{T}$ does not have any exceptional points [Regimes (I), (II), (III), (IV), and (V) in Fig.~\ref{schematic}] and therefore is a diagonalizable matrix. This scenario is summarized in Table.~\ref{table1}.  Therefore one can use Eq.~\ref{matrixexponent} to explicitly write down the elements of $\mathbf{T}^{-N}$ as,
\begin{widetext}
\begin{align}
\label{eq:terms_Tinv}
\left\langle 3 \vert \mathbf{T}^{-N} \vert 3 \right \rangle=\psi_3(\lambda_1)\phi_3(\lambda_1)\lambda_1^{-N}+\psi_3(\lambda_1^{-1})\phi_3(\lambda_1^{-1})\lambda_1^{N}+\psi_3(\lambda_2)\phi_3(\lambda_2)\lambda_2^{-N}+\psi_3(\lambda_2^{-1})\phi_3(\lambda_2^{-1})\lambda_2^{N}, \\ \nonumber
\left\langle 4 \vert \mathbf{T}^{-N} \vert 4 \right \rangle=\psi_4(\lambda_1)\phi_4(\lambda_1)\lambda_1^{-N}+\psi_4(\lambda_1^{-1})\phi_4(\lambda_1^{-1})\lambda_1^{N}+\psi_4(\lambda_2)\phi_4(\lambda_2)\lambda_2^{-N}+\psi_4(\lambda_2^{-1})\phi_4(\lambda_2^{-1})\lambda_2^{N}, \\ \nonumber
\left\langle 3 \vert \mathbf{T}^{-N} \vert 4 \right \rangle=\psi_3(\lambda_1)\phi_4(\lambda_1)\lambda_1^{-N}+\psi_3(\lambda_1^{-1})\phi_4(\lambda_1^{-1})\lambda_1^{N}+\psi_3(\lambda_2)\phi_4(\lambda_2)\lambda_2^{-N}+\psi_3(\lambda_2^{-1})\phi_4(\lambda_2^{-1})\lambda_2^{N}, \\ \nonumber
\left\langle 4 \vert \mathbf{T}^{-N} \vert 3 \right \rangle=\psi_4(\lambda_1)\phi_3(\lambda_1)\lambda_1^{-N}+\psi_4(\lambda_1^{-1})\phi_3(\lambda_1^{-1})\lambda_1^{N}+\psi_4(\lambda_2)\phi_3(\lambda_2)\lambda_2^{-N}+\psi_4(\lambda_2^{-1})\phi_3(\lambda_2^{-1})\lambda_2^{N}.
\end{align}
\end{widetext}
Now, collecting all terms in Eq.~\ref{eq:terms_Tinv} together, the denominator of Eq.~\ref{eqcond1}, takes a form, 
\begin{equation}
A_1 + B_1 \lambda_1^N \lambda_2^N  + C_1  \lambda_1^{-N} \lambda_2^{-N} + D_1   \lambda_1^N \lambda_2^{-N}  + E_1 \lambda_1^{-N} \lambda_2^N,\, 
\label{deno}
\end{equation} 
where all the prefactors ($A_1,B_1,C_1, D_1, E_1$) in front of the eigenvalues $\lambda_i, \lambda_i^{-1}, i=1,2$ are $N$ independent. The subscript $1$ here represents the coefficients associated with the denominator. 
%$A_1,B_1,C_1, D_1, E_1, A_2,B_2, C_2, D_2$ are the $N$-independent constant terms.
Analogously, the numerator can be expressed as,
\begin{equation}
A_2 \lambda_1^N + B_2 \lambda_1^{-N}+ C_2 \lambda_2^N + D_2 \lambda_2^{-N},
\end{equation} 
where once again all the prefactors ($A_2,B_2,C_2, D_2$) are $N$ independent and the subscript $2$ represents the coefficients associated with the numerator. It is important to note that, the expression for the denominator in Eq.~\ref{deno}, terms such as  $\lambda_i^{2N}$  and $\lambda_i^{-2N}, i=1,2$ do not appear and exactly cancel out. In what follows, we now discuss the scaling of $g_{1N}$ with $N$ for different cases corresponding to different values of $\mu$ with no exceptional points. 

\vskip 0.2cm
\noindent
\textit {Below lower band edge} [Regime (I), $\mu<-2t_1-2t_2$]: For $n=2$, the regime below the lower band edge corresponds to $\mu<-2t_1-2t_2$. In Fig.~\ref{schematic} this regime is indicated by the symbol (I).  In this regime, transfer matrix eigenvalues are always real and therefore are of the form $\lambda_1=-e^{\kappa_1}=e^{i\pi}e^{\kappa_1}$ and $\lambda_2=e^{\kappa_2}$, where $\kappa_1, \kappa_2 >0$, and two other eigenvalues being $\lambda_1^{-1}, \lambda_2^{-1}$ (see Table.~\ref{table1}). 
%Considering even number system size $N$, we can write $\lambda_1^N=e^{\kappa_1 N}$. 
Thus, from Eq.~\ref{eqcond1} we get,
\begin{widetext}
\begin{align}
\mathbf{g}_{1N} \sim \frac{A_2 e^{\kappa_1 N} + B_2 e^{-\kappa_1 N}+ C_2 e^{\kappa_2 N} + D_2 e^{-\kappa_2 N}}{A_1 + B_1 e^{(\kappa_1+\kappa_2)N} + C_1 e^{-(\kappa_1+\kappa_2)N}+ D_1 e^{(\kappa_1-\kappa_2)N} + E_1 e^{(-\kappa_1+\kappa_2)N}}.  \end{align}
\end{widetext}
Choosing $\kappa_1, \kappa_2$ such that $\kappa_1>\kappa_2$ and neglecting the exponentially decaying terms in the large $N$ limit, we obtain, 
%and dividing both the denominator and the numerator by $e^{\kappa_1 N}$ and 
\begin{equation}
\mathbf{g}_{1N} \sim \frac{A_2}{B_1 e^{\kappa_2 N}} \sim e^{-\kappa_2 N} 
\end{equation}
which implies
\begin{equation}
|\mathbf{g}_{1N}|^2 \sim e^{-2 \kappa_2 N} \sim \lambda_2^{-2N}.
\end{equation} 
As a result, below the lower band edge, the NESS conductance always decays exponentially with the system size $N$. The corresponding localization length is set by $\kappa_2$ where $\kappa_2$ is related to the smallest eigenvalue $\lambda_2$ of the transfer matrix $\mathbf{T}$ i.e., $\lambda_2 = e^{\kappa_2}$. %Similar analysis can be done with odd system size.
%\textcolor{red}{STOP}
%conductance $\mathcal{G}(\mu)$ in this regime.
\\
\vskip 0.2 cm
\noindent
\textit{Within the band edges [Regime (II) and Regime (III)]:} Let us now discuss the scaling of conductance when the chemical potential $\mu$ is within the band edge. Recall that, the lower band edge always occurs at energy $-2 t_1-2t_2$ (see Fig.~\ref{schematic}). However, the energy corresponding to the upper band edge depends on relative values of $t_1, t_2$ as given in Eq.~\ref{resultn=2}. Thus two distinct regimes [Regime (II) and Regime (III)] emerge within the band edges which is clearly shown in Fig.~\ref{schematic}. 

For regime (II) of Fig.~\ref{schematic} with $-2 t_1 -2 t_2 <\mu < 2 t_1 -2 t_2$, the transfer matrix $\mathbf{T}$ eigenvalues are $\lambda_1=-e^{\kappa_1}$, $\lambda_2=e^{i \kappa_2}$, $\lambda_1^{-1}$ and $\lambda_2^{-1}$, where $\kappa_1, \kappa_2 >0$ (see Table.~\ref{table1}). We therefore obtain from Eq.~\ref{eqcond1},  
\begin{widetext}
\begin{align}
\mathbf{g}_{1N} \sim \frac{A_2 e^{\kappa_1 N} + B_2 e^{-\kappa_1 N}+ C_2 e^{i \kappa_2 N} + D_2 e^{-i \kappa_2 N}}{A_1 + B_1 e^{(\kappa_1+i \kappa_2)N} + C_1 e^{-(\kappa_1+i \kappa_2)N}+ D_1 e^{(\kappa_1-i \kappa_2)N} + E_1 e^{(-\kappa_1+i \kappa_2)N}}.
\label{u-b-e}
\end{align}
\end{widetext}
In the large $N$ limit, Eq.~\ref{u-b-e} simplifies to,
%Again dividing both the numerator and denominator by $e^{\kappa_1 N}$ and neglecting the exponentially decaying terms, 
\begin{equation}
\mathbf{g}_{1N}\sim \frac{A_2}{B_1 e^{i \kappa_2 N}+ D_1 e^{-i \kappa_2 N}}
\end{equation} 
thus implying $|\mathbf{g}_{1N}|^2\sim N^0$ or ballistic transport.  For regime (III) in Fig.~\ref{schematic}, $2 t_1-2 t_2<\mu<2 t_2 + \frac{t_1^2}{4 t_2}$ with $t_2>t_1/4$, the eigenvalues of $\mathbf{T}$ are all complex and given as $\lambda_1=e^{i\kappa_1}$ and $\lambda_2=e^{i \kappa_2}$ (see Table.~\ref{table1}). Thus, all the terms in Eq.~\ref{eqcond1} will have oscillatory dependence on $N$ indicating once again ballistic transport. 

An interesting situation arises for $\mu=2t_1-2t_2$ corresponding to line D in Fig.~\ref{schematic} with $t_2>t_1/4$. Any point along this line corresponds to a second-order exceptional point of $\mathbf{T}$. Nonetheless, despite being an exceptional point, the corresponding NESS conductance is ballistic and this will be elaborate on later. 
%when we describe the conductance of Table.~\ref{table2}.
\vskip 0.2 cm
\noindent
{\it Above the upper band edge [Regime (IV) and Regime (V)]:} 
%We now proceed with the discussion on the scaling of NESS conductance when $\mu$ is above the upper band edge.
Once again depending on the relative values of hopping $t_1$ and $t_2$, two distinct regimes [Regime (IV) and Regime (V)] appear above the upper band edge (see Fig.~\ref{schematic}). Above the upper band edge when $\mu>2t_2 + \frac{t_1^2}{4 t_2}$, it corresponds to the regime (IV) of Fig.~\ref{schematic}. In this case, the eigenvalues of $\mathbf{T}$ are $\lambda_1=-e^{\kappa_1+ i \kappa_2}$ and $\lambda_2=-e^{\kappa_1-i \kappa_2}$ where $\kappa_1, \kappa_2 >0$ (see Table.~\ref{table1}). Now following Eq.~\ref{eqcond1}, we can write,
\begin{widetext}
\begin{align}
\label{u-u}
\mathbf{g}_{1N} \sim \frac{A_2 e^{(\kappa_1+i \kappa_2) N} + B_2 e^{-(\kappa_1+ i \kappa_2) N}+ C_2 e^{(\kappa_1-i \kappa_2) N} + D_2 e^{-(\kappa_1-i \kappa_2) N}}{A_1 + B_1 e^{2\kappa_1N} + C_1 e^{-\kappa_1 N}+ D_1 e^{2i \kappa_2 N} + E_1 e^{-2i \kappa_2N}}.
\end{align}
\end{widetext}
In the large $N$ limit, Eq.~\ref{u-u} reduces to, 
\begin{equation}
 \mathbf{g}_{1N}\sim (A_2 e^{i \kappa_2 N}+ C_2 e^{-i\kappa_2 N})\,e^{-\kappa_1 N}
\end{equation}
implying exponentially decaying transport. 

With $t_2<t_1/4$, $2 t_1-2t_2<\mu< 2 t_2 + t_1^2/4t_2$, corresponds to the above upper band edge i.e., regime (V) of Fig.~\ref{schematic}. Eigenvalues  of transfer matrix $\mathbf{T}$ are $\lambda_1=-e^{\kappa_1}$ and $\lambda_2=-e^{\kappa_2}$, where $\kappa_1, \kappa_2 >0$ (see Table.~\ref{table1}). This is exactly like the situation below the lower band edge i.e., regime (I) of Fig.~\ref{schematic} and therefore shows exponentially decaying conductance $\mathcal{G}(\mu)$ with system size.  For $t_2<t_1/4$, at $\mu=2 t_2 + \frac{t_1^2}{4 t_2}$ corresponds to another interesting situation and is an exceptional line E in Fig.~\ref{schematic}. Nonetheless, despite being an exceptional point, the corresponding NESS conductance is exponentially decaying and this will be elaborated on later. 

To summarize, we have provided a detailed analytical understanding of NESS conductance scaling within and outside the band edges following the transfer matrix eigenspectra that perfectly matches with direct numerics as shown in Fig.~\ref{cond1} and Fig.~\ref{cond3}. In other words, we analytically show the ballistic transport within the band edges and exponentially decaying transport outside the band edges of the lattice system.   Next, we discuss a situation when the transfer matrix $\mathbf{T}$ has exceptional points i.e., along the lines A, B, C, D, and E and the point $\Gamma_{e}$ and is therefore
non-diagonalizable in nature. Recall that this scenario is summarized in Table. \ref{table2}. 
%{\color{red}{STOP}}
\\
\vskip 0.2 cm
\noindent
\noindent
{\textit{At the lower band edge (exceptional line A of Fig.~\ref{schematic}):}} Let us now discuss the NESS conductance scaling with system size at the lower band edge which always occurs at $k=0$ with energy $\mu= -2 t_1-2t_2$, as given in Eq.~\ref{lower-2}. This corresponds to the exceptional line A of Fig.~(\ref{schematic}). Interestingly, for any point on this line, the eigenvalues of $\mathbf{T}$ are given as 
$\lambda_1=1$, $\lambda_2=-e^{\kappa_1}= e^{i \pi}e^{\kappa_1}$ , $\lambda_1^{-1}$, $\lambda_2^{-1}$ where $\kappa_1 >0$ (see Table.~\ref{table2}). Note that $\kappa_1$ in general is a function of $\mu$. Therefore, the lower band edge corresponds to a second-order exceptional line, and hence the transfer matrix $\mathbf{T}$ can be brought to a Jordan normal form $\mathbf{J}$ (see Eq.~\ref{Jordan}). This is given by
%lower band-edge for both $t_2\leq t_1/4$ and $t_2>t_1/4$,  Here, $J$ has the form, 
\begin{equation}
\mathbf{J}=\begin{pmatrix}
 e^{i \pi} e^{\kappa_1} & 0 & 0& 0 \\
 0 &  e^{-i \pi} e^{-\kappa_1} & 0 & 0 \\
 0 & 0 & 1 & 1  \\
 0 & 0 & 0 & 1
\end{pmatrix}
\end{equation} 
and 
\begin{equation}
    \mathbf{J}^{-N} = \begin{pmatrix}
 e^{-\kappa_1 N} e^{-i \pi N} & 0 & 0& 0 \\
 0 & e^{\kappa_1 N} e^{i \pi N}  & 0 & 0 \\
 0 & 0 & 1 & -N  \\
 0 & 0 & 0 & 1
\end{pmatrix}.
\label{J-N}
\end{equation}
In Eq.~\ref{J-N} one of the matrix elements is $-N$ which plays a pivotal role in dictating the scaling of the NESS conductance as we will see now. Using Eq.~\ref{eqcond1}, we can obtain an expression for $\mathbf{g}_{1N}$ as,
\begin{widetext}
\begin{align}
    \textbf{g}_{1N} \sim
 \frac{A_2 + B_2 e^{\kappa_1 N} + C_2 e^{-\kappa_1 N} + D_2 N}{A_1 + B_1 e^{\kappa_1 N}+  C_1 e^{-\kappa_1 N}+   D_1 N e^{\kappa_1 N}+  E_1 N e^{-\kappa_1 N} + F_1 N}.
\end{align}
\end{widetext}
Now, taking the large $N$ limit, we obtain,
\begin{equation}
\mathbf{g}_{1N} \sim \frac{1}{N}. 
\end{equation}
As a result, $|\mathbf{g}_{1N}|^2 \propto {1}/{N^2}$, implying subdiffusive scaling of NESS conductance at the lower band edge. Our analytical findings are rigorously verified by direct numerics as shown in Fig.~\ref{cond3}(a).  

\vskip 0.2 cm
\noindent
{\textit{At the upper band edge (exceptional lines B  and C and exceptional point $\Gamma_e$ of Fig.~\ref{schematic}):}} We now discuss the conductance scaling when the chemical potential $\mu$ is located at the upper band edge. This band edge is comprised of three parts: exceptional line B, exceptional line C, and exceptional point $\Gamma_e$. 
 Let us discuss NESS scaling for each of these cases separately. 
 For the case $t_2< t_1/4 $, the upper band edge is located at $k=\pi$ with corresponding energy $\mu=2 t_1 - 2t_2$. This corresponds to line B of the phase diagram Fig.~\ref{schematic}. In this case, the eigenvalues of $\mathbf{T}$ are $\lambda_1=-1$, $\lambda_2= e^{i \pi} e^{\kappa_1}$, $\lambda_1^{-1}$, and $\lambda_2^{-1}$ and hence once again it's a second-order exceptional point (see Table.~\ref{table2}). Here 
 $\kappa_1 >0$. Therefore, exactly like the lower band edge case, one can show that the NESS conductance scales subdiffusively as $1/N^2$. This is also clearly shown in Fig.~\ref{cond3}(b). 

When $t_2>t_1/4$, the location of the upper band edge does not occur at $k=\pi$. The corresponding energy is $\mu=2t_2+t_1^2/4 t_2$ which is represented by line C in Fig.~\ref{schematic}. Interestingly, in this scenario, the eigenvalues of $\mathbf{T}$ are $\lambda_1=e^{i \kappa_1}$ and $\lambda_2=e^{-i \kappa_1}$, $\lambda_1^{-1}$, and $\lambda_2^{-1}$ (see Table~\ref{table2}). Here $\kappa_1 >0$. As a result, the four eigenvalues form  two complex conjugate pairs of two each. This implies that there are two second-order exceptional points that are complex, unlike the case when the upper band edge is located at $k=\pi$ i.e., line B.
Thus, once again the transfer matrix $\mathbf{T}$ can be brought to a Jordan-normal form given as,
\begin{equation}
\mathbf{J}=\begin{pmatrix}
 e^{i \kappa_1} & 1 & 0& 0 \\
 0 & e^{i\kappa_1} & 0 & 0 \\
 0 & 0 & e^{-i\kappa_1} & 1  \\
 0 & 0 & 0 & e^{-i\kappa_1}
\end{pmatrix}
\end{equation}
and 
\begin{equation}
\mathbf{J}^{-N}=
\begin{pmatrix}
 e^{-i \kappa_1 N} & -N e^{-i \kappa_1 (N+1)} & 0& 0 \\
 0 & e^{-i \kappa_1 N} & 0 & 0 \\
 0 & 0 &  e^{i \kappa_1 N} & -N e^{i \kappa_1 (N+1)} \\
 0 & 0 & 0 &  e^{i \kappa_1 N}
\end{pmatrix}.
\end{equation}
%Thus, the $N$-dependent terms in numerator is $ B_2 e^{-i \kappa_1 N}+ C_2 e^{i \kappa_1 N}+ D_2 N e^{-i \kappa_1 (N+1)}+ E_2 N e^{i \kappa_1 (N+1)} $. The denominators will have the terms, $A_1+ B_1 N^2 + C_1 N e^{-i \kappa_1}+ D_1 N e^{i \kappa_1}+ E_1 e^{i \kappa_1(2 N+1)}+ F_1 e^{-i \kappa_1 (2N+1)}$. 
With this result, Eq.~\ref{eqcond1} can be written as,
\begin{widetext}
\begin{align}
\label{upperbandedge}
\mathbf{g}_{1N}\sim
 \frac{B_2 e^{-i \kappa_1 N}+ C_2 e^{i \kappa_1 N}+ D_2 N e^{-i \kappa_1 (N+1)}+ E_2 N e^{i \kappa_1 (N+1)}}{A_1+ B_1 N^2 + C_1 N e^{-i \kappa_1}+ D_1 N e^{i \kappa_1}+ E_1 e^{i \kappa_1(2 N+1)}+ F_1 e^{-i \kappa_1 (2N+1)}}.
\end{align}
\end{widetext}
In the large-$N$ limit, Eq.~\ref{upperbandedge} simplifies to, 
\begin{equation}
  \mathbf{g}_{1N}\sim \frac{D_2 e^{-i \kappa_1 N}+  E_2 e^{i\kappa_1 N}}{B_1 N}
\end{equation}
and as a result, the NESS conductance shows interesting oscillations set by $\kappa_1$ along with overall $1/N^2$ subdiffusive scaling. 
This is another central finding of this paper. Our analytical results have been corroborated with the direct numerical simulations shown in Fig.~\ref{cond3}(c). 

Let us now discuss the conductance scaling at the exceptional point $\Gamma_e$ in Fig.~\ref{schematic}. This special point occurs for $t_2=t_1/4$ and corresponds to the upper band edge energy $\mu= 2t_1 - 2t_2$. At this special point, all four eigenvalues and eigenvectors of $\mathbf{T}$ coalesce (see Table.~\ref{table2}) thereby yielding a fourth-order exceptional point. The corresponding Jordan-Normal form at this fourth-order exceptional point $\Gamma_e$ is given by
 \begin{equation}
     \mathbf{J}=
 \begin{pmatrix}
-1 & 1 & 0 & 0 \\
0 & -1 & 1 & 0 \\
0 & 0 & -1 & 1 \\
0 & 0 & 0 & -1
\end{pmatrix}
\end{equation}
and 

\begin{widetext}
\begin{equation}
 \mathbf{J}^{-N}=
 \begin{pmatrix}
 e^{-i\pi N} & N  e^{-i\pi N} &  e^{-i\pi N}\frac{1}{2} N(N+1) &  e^{-i\pi N}\frac{1}{6} N(N+1)(N+2) \\
 0 &  e^{-i\pi N} & N  e^{-i\pi N} &  e^{-i\pi N} \frac{1}{2} N(N+1) \\
 0 & 0 &  e^{-i\pi N} & N  e^{-i\pi N} \\
 0 & 0 & 0 &  e^{-i\pi N}
\end{pmatrix}.
\label{high-order}
\end{equation}
\end{widetext}
It is interesting to note that, the matrix elements $\mathbf{J}^{-N}$ in Eq.~\ref{high-order} contain terms up to $O(N^3)$ which is in stark contrast with all the other cases where exceptional points were of second order. 
However, the final system-size scaling of NESS conductance still shows $1/N^2$ subdiffusive scaling and therefore extremely robust against the order of the exceptional points of transfer matrices indicating a strong presence of universality. Below we provide the details.  
In this case, to determine the scaling of conductance, we need to know the explicit form of the transformation matrix $
\mathbf{R}$, as defined in Eq.~\ref{Jordan}. We obtain, 
\begin{equation}
 \mathbf{R}=\begin{pmatrix}
 0 & 0 & 0 & 1 \\
 0 & 0 & 1 & 1 \\
 0 & 1 & 2 & 1\\
 1 & 3 & 3 & 1
\end{pmatrix}.   
\end{equation}
 Thus, using this form of $\mathbf{R}$, we obtain the different matrix elements for $\mathbf{T}$ as,
 \begin{eqnarray}
 \left\langle 3 \vert T^{-N} \vert 3 \right \rangle &=&\frac{e^{-i\pi N}}{2} (2+N-2N^2-N^3), \nonumber \\
 \left\langle 3 \vert T^{-N} \vert 4 \right \rangle &=& -\frac{e^{-i\pi N}}{6} N (N+1)(N+2), \nonumber \\ 
 \left\langle 4 \vert T^{-N} \vert 3 \right \rangle &=& \frac{e^{-i\pi N}}{2} N(N+1)(N+3), \nonumber \\
 \left\langle 4 \vert T^{-N} \vert 4 \right \rangle &=& \frac{e^{-i\pi N}}{6} (6+11N+6N^2+N^3). 
 \label{group-eq}
 \end{eqnarray}
Substituting the expressions obtained in Eq.~\ref{group-eq} in Eq.~\ref{eqcond1}, we receive,
\begin{align}
\mathbf{g}_{1N}\sim
\frac{A_1 N^3 + B_1 N^2 + C_1 N + D_1}{A_2 N^4 + B_2 N^3+ C_2 N^2 + D_2 N + E_2}    
\end{align}
which in the large $N$ limit gives $\mathbf{g}_{1N}\sim 1/N$. Thus, conductance scales as $1/N^2$ like the other band edges i.e., exceptional lines A, B, and C even though the transfer matrix has a higher-order exceptional point. This analysis also matches with our numerical findings as shown in Fig.~\ref{cond3}(f). 

\vskip 0.2 cm
\noindent
{\textit{Within the band edge, (along the exceptional line D of Fig.~\ref{schematic}):}} The exceptional line D emerges for  $t_2>t_1/4$ which separates regime (II) and regime (III) of Fig.~\ref{schematic}. This line always occurs within the two band edges at $\mu=2t_1-2t_2$. The transfer matrix eigenvalues in this case are given as $\lambda_1=-1$, $\lambda_2=e^{i\kappa_1}$, $\lambda_1^{-1}$ and $\lambda_2^{-1}$ (see Table.~\ref{table2}). Here $\kappa_1>0$. As a result, once again the transfer matrix $\mathbf{T}$ is not diagonalizable and can be brought to a Jordan normal form  given by, 
\begin{align}
\mathbf{J}=\begin{pmatrix}
 e^{i\kappa_1} & 0 & 0& 0 \\
 0 & e^{-i\kappa_1} & 0 & 0 \\
 0 & 0 & -1 & 1  \\
 0 & 0 & 0 &- 1
\end{pmatrix}
\end{align}
and $\mathbf{J}^{-N}$ is given by,
\begin{align}
\mathbf{J}^{-N}=\begin{pmatrix}
 e^{-i\kappa_1 N} & 0 & 0& 0 \\
 0 & e^{i \kappa_1 N} & 0 & 0 \\
 0 & 0 & e^{-i \pi N} & N e^{-i \pi N}  \\
 0 & 0 & 0 & e^{-i\pi N}
\end{pmatrix}.
\end{align}
%Note that the similarity transformation matrix $\mathbf{R}$, as defined in Eq.~\ref{Jordan}, does not depend on the system size $N$. 
As a result, following Eq.~\ref{eqcond1} we obtain, 
\begin{widetext}
\begin{align}
    \mathbf{g}_{1N} \sim 
 \frac{A_2 + B_2 e^{i\kappa_1 N} + C_2 e^{-i \kappa_1 N} + D_2 N}{A_1 + B_1 e^{i \kappa_1 N}+  C_1 e^{-i \kappa_1 N}+   D_1 N e^{i \kappa_1 N}+  E_1 N e^{-i \kappa_1 N} + F_1 N}.
 \label{u-b-e-e}
\end{align}
\end{widetext}
In the large $N$ limit, Eq.~\ref{u-b-e-e} simplifies to 
\begin{equation}
  \mathbf{g}_{1N} \sim \frac{ D_2}{D_1  e^{i \kappa_1 N}+  E_1  e^{-i \kappa_1 N} + F_1 }
\end{equation}
which produces ballistic transport and is further supported by direct numerics and shown in Fig.~\ref{cond3}(d). It is worth noting that this ballistic behavior occurs even in presence of exceptional points. This further implies that albeit the points are exceptional in nature, the fact that they appear within the band edge causes ballistic transport. 

\vskip 0.2 cm
\noindent
{\textit{\textit Above the band edge (along the exceptional line E of Fig.~\ref{schematic}):}} The exceptional line E emerges for $t_2<t_1/4$ separating regime (IV) and (V) of Fig.~\ref{schematic}. This line always occurs above the upper band edge and corresponds to energy $\mu=2 t_2 + t_1^2/4t_2$. The transfer matrix eigenvalues are 
 $\lambda_1=e^{i \pi} e^{\kappa_1}$ and $\lambda_2= e^{i \pi}  e^{-\kappa_1}$ where $\kappa_1>0$ (see Table.~\ref{table2}). Interestingly, here the transfer matrix has a pair of second-order exceptional points. The Jordan-normal form is given by,
\begin{align}
\mathbf{J}=\begin{pmatrix}
 e^{i \pi} e^{ \kappa_1} & 1 & 0& 0 \\
 0 & e^{i \pi} e^{\kappa_1} & 0 & 0 \\
 0 & 0 & e^{-i \pi} e^{-\kappa_1} & 1  \\
 0 & 0 & 0 & e^{-i \pi} e^{-\kappa_1}
\end{pmatrix}
\end{align}
and
\begin{widetext}
    \begin{align}
\mathbf{J}^{-N}=\begin{pmatrix}
  e^{-i \pi N} e^{- \kappa_1 N} & -N e^{- (i \pi + \kappa_1) (N+1)} & 0& 0 \\
 0 & e^{-i \pi N} e^{- \kappa_1 N} & 0 & 0 \\
 0 & 0 & e^{i \pi N} e^{ \kappa_1 N} & -N e^{ (i \pi+\kappa_1) (N+1)} \\
 0 & 0 & 0 & e^{i \pi N} e^{ \kappa_1 N}
\end{pmatrix}.
\end{align}
%\end{widetext}
With that, Eq.~\ref{eqcond1} can be written as,
%\begin{widetext}
\begin{align}
\label{upperbandedge1}
\mathbf{g}_{1N}\sim \frac{ B_2 e^{- \kappa_1 N}+ C_2 e^{ \kappa_1 N}+ D_2 N e^{- \kappa_1 (N+1)}+ E_2 N e^{ \kappa_1 (N+1)}}{A_1+ B_1 N^2 + C_1 N e^{- \kappa_1}+ D_1 N e^{ \kappa_1}+ E_1 e^{ \kappa_1(2 N+1)}+ F_1 e^{- \kappa_1 (2N+1)}}
\end{align}
\end{widetext}
which in the large $N$ limit gives $\mathbf{g}_{1N}\sim e^{-\kappa_1 N}$. Thus conductance shows exponentially decaying scaling with system size which also matches with the direct numerics as shown in Fig.~\ref{cond3}(e). Note that, albeit the points along line E are exceptional in nature, the fact that they appear outside the band edge causes exponentially suppressed transport.

%\textcolor{red}{STOP}

Next in section.~\ref{general-sec}, we comment on the NESS conductance scaling for general finite-range hopping systems.

\subsection{Comment on general finite-range hopping system}
\label{general-sec}
It is possible to generalize the study performed in Sec.~\ref{n-2-sec} for any finite range hopping systems, i.e., $n=3, 4, 5, \cdots$. Accordingly, following Eq.~\ref{discretescro2} the $2n \times 2n$ transfer matrix $\mathbf{T}(\mu)$ can be constructed and its eigenspectra can be subsequently analyzed. The NESS conductance and its system size scaling behavior can then be addressed using Eqs.~\ref{rescaledg} and \ref{inverse5}. We note that there is a general framework obeyed by all finite range models irrespective of the range of hopping parameter $n$ which we elaborate on below. Without loss of generality, we again set $t_1=1$.

\begin{figure*}
\includegraphics[width=2\columnwidth]{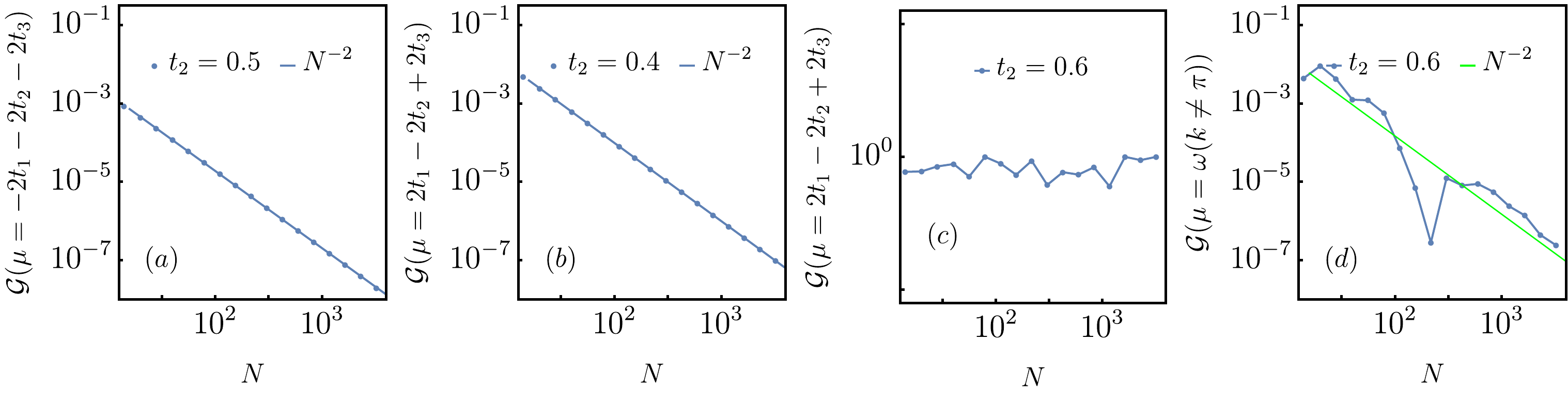} 
\caption{Plot for system size scaling of NESS conductance for $n=3$ at various exceptional points of the exceptional hyper-surfaces. In all these figures we set $t_1=1$, $t_3=1/9$, and the value of $t_2$ are displayed in the corresponding plots. (a) Subdiffusive $1/N^2$ scaling is reported at the lower band edge that corresponds to $k=0$, (b) subdiffusive  $1/N^2$ scaling at $k=\pi$ that corresponds to the upper band edge, (c) ballistic $N^{0}$ scaling with system size at $k= \pi$ which does not correspond to the upper band edge, and (d) oscillatory behaviour with an overall envelope that is subdiffusive in nature with $1/N^2$ scaling. This occurs at the upper band edge with $k \approx 1.92$ (i.e., $k \neq \pi$).}
\label{cond4} 
\end{figure*}

\begin{enumerate}
    \item At $k=0$, the transfer matrix $\mathbf{T}(\mu)$ needs to be evaluated at the lower band edge energy $\mu= - 2 \sum_{m=1}^{n} t_m$ (see Eq.~\ref{lower band edge}). This naturally defines a $(n-1)$ dimensional hyper-surface and when $\mathbf{T}(\mu)$ is evaluated at any point on this hyper-surface it will have an exceptional point. This can be understood as follows: at the lower band edge, $\theta=0$  where recall that $\theta$ is related to the eigenvalue of $\mathbf{T}(\mu)$ by $\lambda=e^{i \theta}$ (see Eq.~\ref{l-theta}) is always a solution of Eq.~\ref{Ftheta}. Thus at least two eigenvalues with value $1$ and corresponding eigenvectors of $\mathbf{T}(\mu)$ coalesce and hence an exceptional point.  It then turns out that the corresponding NESS conductance is subdiffusive with universal $1/N^2$ scaling, irrespective of the value $n$. We illustrate this for the case $n=3$ in Fig.~\ref{cond4} (a). 

    \item At $k=\pi$, $\mathbf{T}(\mu)$ needs to be evaluated at $\mu=\omega(k=\pi)$ following Eq.~\ref{upper band edge}. This once again forms a $(n-1)$ dimensional hyper-surface and if $\mathbf{T}(\mu)$ is evaluated at any point on this hyper-surface, it will have an exceptional point. However, interestingly this point may not always correspond to the upper band edge. This, therefore, yields  two different scenarios: When $k=\pi$ corresponds to the usual upper band edge we obtain subdiffusive scaling $1/N^2$ for NESS conductance for reasons similar to that for $n=2$ case. We illustrate this for the case $n=3$ in Fig.~\ref{cond4} (b). In contrast, when $k=\pi$ does not correspond to the upper band edge, it naturally implies that $k=\pi$ point is located inside the band edges. Therefore albeit being an exceptional point, the NESS conductance will show ballistic behavior $N^{0}$ and this is illustrated explicitly for the $n=3$ case in Fig.~\ref{cond4} (c). In the scenario when the upper band edge is located at some other value of $k \neq \pi$, the eigenvalues of $\mathbf{T}(\mu)$ come in complex conjugate pairs and are exceptional in nature. This gives rise to an oscillatory behavior with an overall envelope of $1/N^2$ scaling for NESS conductance. This is illustrated in Fig.~\ref{cond4} (d).

    \item Similar to $n=2$, exceptional hyper-surface may likely emerge above the upper band edge. However, the transfer matrix  $\mathbf{T}(\mu)$ evaluated at points on this hyper-surface have real eigenvalues (not equal to 1) and are exceptional in nature. The resulting conductance will be exponentially suppressed with system size. 
    %This is illustrated in Fig.~\ref{cond4} (e). 
    
    \item  It is worth mentioning that, fourth-order exceptional points will always emerge at $k=\pi$ when the hopping strengths satisfy the condition in Eq.~\ref{condition3}. For $n \geq 2$, the fourth order exceptional point $\Gamma_e$ will become a $(n-2)$ hyper-surface of fourth order exceptional points.  With odd $n$, in presence of such fourth-order exceptional points in the transfer matrix, NESS conductance will always show ballistic behaviour as $k=\pi$ does not correspond to the upper band edge as discussed in Eq.~\ref{high-der}. Whereas with even $n$, at $k=\pi$ the NESS conductance will show subdiffusive transport with scaling $1/N^2$ as it corresponds to the upper band edge. 
    
    %There is a  possibility that $2n$-th order exceptional point may appear for finite range lattice with hopping range $n$. Needless to mention, we find that searching for such higher-order exceptional points for $n>2$ is both analytically and numerically highly challenging, given the high dimensionality of the parameter space. 
    %Also, because of the sensitivity of the exceptional point, without knowing the exact conditions, it is difficult to locate the exceptional point numerically.  
    
    %Nonetheless one can perform a numerical search for the optimal values of hoppings where the transfer matrix has $2n-$th order exceptional point. 
    
   % It is worth mentioning that, fourth-order exceptional points for any range of hopping $n$, are relatively feasible to obtain. 
\end{enumerate}
Having established a strong sense of universality in NESS transport properties with respect to range of the hopping parameter, an important question is that of robustness to imperfections in realistic systems. In Section~\ref{robust} we address this point.

\begin{figure*}
\includegraphics[width=2\columnwidth]{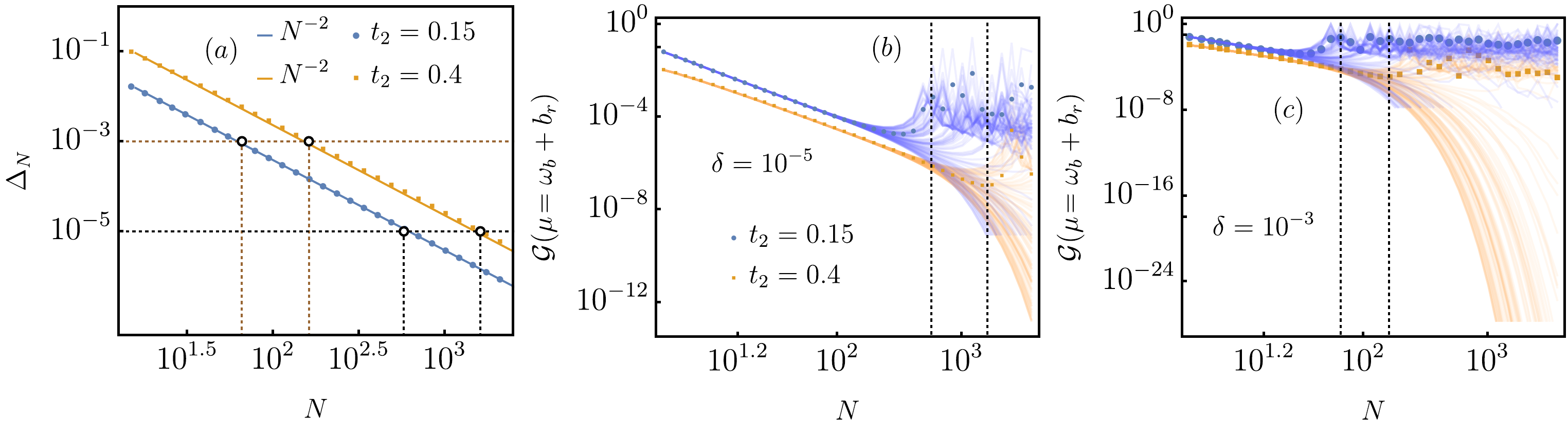} 
\caption{Figures supporting the robustness of our finding of anomalous transport to various kinds of disorder. [Left panel] The figure shows the gap $\Delta_N$ (for a clean system, i.e., $\epsilon_i= 0 $ in Eq.~\ref{dis-on}) between the upper or lower (blue or orange, respectively) band energy of a finite size system (i.e., finite $N$) to that of its corresponding thermodynamic limit (i.e., $N \to \infty$). The gap $\Delta_N$ for a clean system decreases with system size $N$ as $1/N^2$. This figure plays a pivotal role in getting an estimate of allowed disorder that does not destroy anomalous transport. This estimate is made as follows: The vertical black and brown dotted lines represent four pairs $(\Delta_N^{*}, N^{*})$ shown by four black circles. The right element of each pair $(\Delta_N^{*}, N^{*})$ gives an estimate of $N^{*}$ up to which anomalous transport is robust. The left element of each pair $(\Delta_N^{*}, N^{*})$ gives the corresponding estimate for the allowed disorder strength i.e., $\delta \sim \Delta_N{*}$. (b) [Middle panel] The figure shows the robustness of the anomalous transport in presence of weak disorder $\delta=10^{-5}$ both in on-site potential in Eq.~\ref{dis-on} and in the chemical potential $\mu = \omega_b + b_r$ where $b_r$ is a random number chosen from a uniform distribution $[-\delta,\delta]$. We have plotted the conductance for 100 different realizations (represented by light color lines) as well as the mean value (represented by dots). The deviation from anomalous behavior occurs approximately near $N^{*}$ (represented by dotted vertical lines) that is estimated using the plot in the left panel of the same figure. Likewise, the left panel of the same figure gives an estimate of $\delta$ which in this case is $\delta= 10^{-5}$. [Right panel] Figure similar to the middle panel for the pairs $(\Delta_N^{*}, N^{*})$ which yield $\delta=10^{-3}$. Note that a similar analysis will hold even when similar disorder is present in the hopping parameters $t_1$ and $t_2$.} 
\label{robustness} 
\end{figure*}

\subsection{Robustness}
\label{robust}
In this section, we discuss the fate of anomalous transport that occurs at the band edges of the clean lattice system at zero temperature with respect to (i) weak disorder within the system,  (ii) chemical potential fixed near the band edge energies, and (iii) finite but low temperature. Let us first discuss the situation when the clean lattice Hamiltonian $\hat{H}$ in  Eq.~\ref{model} is subjected to weak onsite disorder of strength $\delta$. The Hamiltonian for such a disordered system takes the form, 
\begin{equation}
    \hat{H}= \hat{H} + \hat{H}_{\rm{d}}
\end{equation}
where $\hat{H}_{\rm{d}}$ describes the onsite disorder given as,
\begin{equation}
\hat{H}_{\rm{d}}= \sum_{i=1}^{N} \epsilon_i \, \hat{c}_i^{\dagger} c_i
\label{dis-on}
\end{equation}
with $\epsilon_i$ chosen randomly from a uniform distribution $[0, \delta]$. In presence of such onsite disorder system, we investigate the fate of subdiffusive behaviour. Additionally, we also shift the chemical potential across the band edge by an amount $b_r$ where $b_r$ is a random number chosen from a uniform distribution $[-\delta,\delta]$. In other words, we set 
\begin{equation}
    \mu=\omega_b+b_r.
\label{eq:mu_dis}    
\end{equation}

Before discussing the fate of anomalous transport when subject to disorder we will analyse the gap $\Delta_N$ between the upper or lower band energy of a finite size system (i.e., finite $N$) to that of its corresponding thermodynamic limit (i.e., $N \to \infty$) for a clean system, i.e., $\epsilon_i= 0 $ in Eq.~\ref{dis-on}. In Fig.~\ref{robustness} (a), we plot this gap parameter $\Delta_N$ with system size $N$. We see that $\Delta_N$ decays as $N^{-2}$. For a chosen value of $N^{*}$, there is a $\Delta_{N^*}$ shown by the black circles in Fig.~\ref{robustness} (a). This $\Delta_{N^*}$ provides an estimate for disorder strength $\delta$ for which anomalous scaling is observed approximately up to system size $N^{*}$. In Fig.~\ref{robustness} (b) and (c), we display the robustness of subdiffusive $1/N^2$ scaling with respect to weak disorder $\delta=10^{-5}$ [Fig.~\ref{robustness} (b)] and $\delta=10^{-3}$ [Fig.~\ref{robustness} (c)] by plotting 100 different disorder realizations as well as their disorder averaged values. Note that each disorder realization stands for a particular chemical potential $\mu$ (Eq.~\ref{eq:mu_dis}) and onsite energy $\epsilon_i$ (Eq.~\ref{dis-on}). 
The deviation from the subdiffusive behaviour occurs near a critical finite system size $N^{*}$ which can be clearly seen from Fig.~\ref{robustness}(b) and (c). 

We end this section by making a comment regarding finite (but low) temperature. 
The robustness of the anomalous transport (up to a critical finite system size) despite having a window around the band edge as per Eq.~\ref{eq:mu_dis} also suggests a possible window of temperature (albeit small) where one does not destroy the subdiffusive nature of scaling. We expect subdiffusive behaviour of conductance up to a inverse temperature $\beta~\delta>1$ or $\beta \Delta_{N^{*}} >1$.

From the above detailed analysis, one can conclude that the NESS scaling of conductance with system size at both the band edges with or without oscillations remains robust in presence of (i) weak on-site disorder, (ii) fine-tuned energies across the band edges, and (iii) low temperatures.

\section{Summary and outlook}
\label{sec:summary}
In summary, we have performed a detailed analysis of the non-Hermitian properties of transfer matrices and exceptional hyper-surfaces and their impact on the scaling of NESS conductance for arbitrary finite range hopping model (Table~\ref{table1} and Table~\ref{table2}). We have established the connection between non-equilibrium steady state (NESS) conductance and underlying non-Hermitian transfer matrix for these lattice models.  We unravel the non-trivial role played 
by exceptional points in determining the universal system size scaling of NESS conductance at the band edges (Table~\ref{table2}, Fig.~\ref{cond3}, Fig.~\ref{cond4}). We further provide evidence that the value of the scaling exponent is remarkably robust to the order of the exceptional point. 
The signature of the upper band edge not being located at $k=\pi$ shows up in the conductance as an interesting oscillation with overall $N^{-2}$ envelope.  It is interesting to note that though the exceptional points appear at very specific energies (and therefore sensitive), none-the-less, the NESS conductance is robust (Fig.~\ref{robustness}) against weak onsite energy and small shift in chemical potential and temperature.  Though, the entire analysis has been done for the linear response regime looking at the conductance behaviour, in the non-linear response regime also, NESS current shows similar non-analytic change (exponentially decaying with system size to ballistic behaviour via subdiffusive transport at the band edges) at one of the band edges. Needless to mention, understanding the microscopic origin of anomalous transport is far from being well understood. The non-Hermitian properties of transfer matrix provides a transparent approach towards understanding the emergence of anomalous scaling of conductance at the band edges and is therefore of extreme relevance to understand the physics of quantum transport.

% We observed that at the band edges the conductance shows universal subdiffusive scaling with system size $1/N^2$ beyond nearest neighbour hopping model also. Previously, we found the universal subdiffusive scaling for long-range hopping model. Thus, this analysis bridges the gap between nearest neighbour hopping and long-range hopping model and also points towards the superuniversility of the $1/N^2$ subdiffusive scaling of conductance. For finite-range hopping model with $n>1$, though the lower band edge is always located at $k=0$ but  the upper band edge can be located other than at $k=\pi$ depending on the conditions on different hopping strengths. Whenever the upper band edge is not located at $k=\pi$, the subdiffusive scaling with system size in conductance features interesting oscillations. This difference crucially depends on the nature of eigenvalues of the exceptional points. Also finite-range hopping model with $n>1$ can give rise to additional exceptional points of transfer matrices which are not at the band edges and these exceptional points do not give rise to subdiffusive behaviour. Most interestingly, with $n>1$, we find fourth order exceptional points of transfer matrices at the upper band edge. These interesting behaviours emerge from the inclusion of additional hoppings with proper conditions but missing in the nearest neighbour hopping model. 

Having done a detailed investigation on the consequence of non-Hermitian transfer matrices and exceptional points in NESS conductance, a natural interesting question is the role of external perturbations such as B{\"u}ttiker voltage probes \cite{Pastawski,self-2-Abhishek,segal-probe-3,buttikerdephase,saha_buttiker,saha-superballistic1} which models incoherent processes within the system without directly taking part in the transport process.
%It is well-known that probes mimic effects due to incoherent processes that a from surrounding environments. 
%Such environments which make transport incoherent without directly taking part in the transport process are modelled by B{\"u}ttiker voltage probes \cite{}. 
Such studies are especially fascinating because usually exceptional points are sensitive to external perturbations and the sensitivity crucially depends on the order of the exceptional points. Hence, it is an interesting and challenging task to see the impact of higher-order exceptional points (that were reported in this work) on conductance due to incoherent processes induced by such probes. Another fascinating and challenging problem is investigating anomalous transport at such exceptional points starting from a many-body interacting Hamiltonian with finite range hopping.

%\textbf{Usually exceptional points are sensitive to external perturbations and the sensitivity crucially depends on the order of the exceptional points. Here,  conductance shows a usual universal subdiffusive scaling $1/N^2$ with system size in presence of higher order exceptional point of transfer matrix. Thus, it will be interesting to see the impact of higher-order exceptional points on conductance due to the external perturbations like environmental effects.}

%It will be interesting to analyze the impact of higher-order exceptional points on conductance upon impact of Buttiker probes.

\section*{Acknowledgement} The authors would like to acknowledge Archak Purkayastha for numerous useful discussions. 
M.~S acknowledges funding from National Postdoctoral Fellowship Scheme (NPDF), SERB file No. PDF/2020/000992. BKA acknowledges the MATRICS grant MTR/2020/000472 from SERB, Government of India and the Shastri Indo-Canadian Institute for providing financial support for this research work in the form of a Shastri Institutional Collaborative
Research Grant (SICRG). B. K. A. would also like to acknowledge funding from National Mission on Interdisciplinary  Cyber-Physical  Systems (NM-ICPS)  of the Department of Science and Technology,  Govt. Of  India through the I-HUB  Quantum  Technology  Foundation, Pune  India. M.K. would like to acknowledge support from the project 6004-1 of the Indo-French Centre for the Promotion of Advanced Research (IFCPAR), Ramanujan Fellowship (SB/S2/RJN-114/2016), SERB Early Career Research Award (ECR/2018/002085) and SERB Matrics Grant (MTR/2019/001101) from the Science and Engineering Research Board (SERB), Department of Science and Technology (DST), Government of India. M.K. acknowledges support of the Department of Atomic Energy, Government of India, under Project No. 19P1112R\&D.\\
\\
\\

\appendix

\section{Details about Bare Green's function for the finite-range hopping model}
\label{sec:appA}
%\subsubsection{Transfer matrix for general finite-range hopping model}
In this section, we provide the details about calculating the Bare Green's function $\mathbf{g}(\mu)$, given in Eq.~\ref{rescaledg}, for a finite range lattice model with system size $N$ and  range of hopping $n$. The details of this calculation can be found in Ref.~\onlinecite{matrix_inversion}. Here we summarize the main points of the derivation to obtain $\mathbf{g}(\mu)$. The calculation of $\mathbf{g}(\mu)$ involves calculating the inverse of $\mathbf{\textbf{M}}(\mu)$, as defined in Eq.~\ref{M-mat}. To obtain the inverse, we  use the identity, $\mathbf{\textbf{M}}(\mu) \mathbf{\textbf{M}}(\mu)^{-1}=\mathbb{I}$ which gives,
\begin{align}
\label{inverse1a}
\sum\limits_{j=1}^N \left\langle i\vert\mathbf{\textbf{M}}(\mu)\vert j\right\rangle \left\langle j\vert\mathbf{\textbf{M}}(\mu)^{-1}\vert k \right\rangle=\delta_{i,k}
\end{align}
Now using Eq.~\ref{elementM_mu}, we can write Eq.~\ref{inverse1a} as,
\begin{align}
\label{inverse1aa}
 \sum\limits_{m=\alpha(i)}^{\eta(i)} a(|m|) \left\langle i+m \vert\mathbf{\textbf{M}}(\mu)^{-1}\vert k \right\rangle&=\delta_{i,k},
\end{align}
where the sum runs from $\alpha(i)=\textrm{Max}\{1-i,-n\}$ and $\eta(i)=\textrm{Min}\{N-i,n\}$. We now define a vector $\mathbf{V}_i(j)$ with dimension $2n \times 1$ given as,
\begin{align}
\mathbf{V}_i(j)=\begin{pmatrix} \left\langle i-n+1 \vert\textbf{M}(\mu)^{-1}\vert j \right\rangle \\
\left\langle i-n+2 \vert\mathbf{\textbf{M}}(\mu)^{-1}\vert j \right\rangle\\ \left\langle i-n+3 \vert\textbf{M}(\mu)^{-1}\vert j \right\rangle\\ \vdots \\ \left\langle i+n \vert\textbf{M}(\mu)^{-1}\vert j \right\rangle \end{pmatrix}.
\label{v-vec}
\end{align}
 Here, $1\leq i,j\leq N$, otherwise $\left\langle i \vert\textbf{M}(\mu)^{-1}\vert j \right\rangle$ is zero. Using the transfer matrix $\mathbf{T}(\mu)$ of dimension $2n\times 2n$ in Eq.~\ref{general_transfer}, we can rewrite Eq.~\ref{inverse1aa} as,
\begin{align}
\label{inverse2a}
\mathbf{T}(\mu) \mathbf{V}_i (j)=\mathbf{V}_{i-1}(j) -\delta_{i,j} \mathbb{I}\vert \mathbf{1} \rangle,
\end{align}
where $\vert \mathbf{1} \rangle$ is a column matrix of dimension $2n \times 1$ with the first element $1$ and all other elements are $0$. By iterating Eq.~\ref{inverse2a}, we can obtain,
\begin{align}
\label{inverse3a}
\mathbf{V}_i (j)= 
\begin{cases}
    \mathbf{T}(\mu)^{-i} \mathbf{V}_{\text{0}} (j),& \text{if }~ j>i\\
    \mathbf{T}(\mu)^{-i} \mathbf{V}_{\text{0}} (j) - \mathbf{T}(\mu)^{-(i-j+1)} \vert \mathbf{1} \rangle          & \text{if} ~j\leq i
\end{cases}
\end{align}
Note that the vector $\mathbf{V}_{\text{0}} (j)$ in Eq.~\ref{inverse3a} contains zeros in its first $n$ elements. Therefore using Eq.~\ref{v-vec} and Eq.~\ref{inverse3a} we finally obtain,
\begin{widetext}
\begin{align}
\label{inverse4a}
\left\langle i \vert\textbf{M}(\mu)^{-1}\vert j \right\rangle= 
\begin{cases}
   \sum\limits_{m=1}^n \left\langle n \vert \mathbf{T}(\mu)^{-i} \vert n+m \right \rangle \left\langle m\vert\textbf{M}(\mu)^{-1}\vert j \right\rangle,& \text{if }~ j>i\\
    \sum\limits_{m=1}^n \left\langle n \vert \mathbf{T}(\mu)^{-i} \vert n+m \right \rangle \left\langle m\vert\textbf{M}(\mu)^{-1}\vert j \right\rangle- \left\langle n \vert \mathbf{T}(\mu)^{-(i-j+1)} \vert 1 \right\rangle          & \text{if} ~j\leq i.
\end{cases}
\end{align}
Eq.~\ref{inverse4a} shows that any matrix element of $\textbf{M}(\mu)^{-1}$ involves the information of $\langle m|\textbf{M}(\mu)^{-1}|j\rangle$ with $m=1,2 \ldots n$. These matrix elements can be determined by noting that the vector $\mathbf{V}_N (j)$ has zeros in its last $n$ elements. Using this fact in Eq.~\ref{inverse4a}, we obtain,
\begin{align}
\label{inverse5a}
 \sum\limits_{m=1}^n \left\langle s+n \vert \mathbf{T}(\mu)^{-N} \vert n+m \right \rangle \left\langle m\vert\textbf{M}(\mu)^{-1}\vert j \right\rangle 
 -\left\langle s+n \vert \mathbf{T}(\mu)^{-(N-j+1)} \vert 1 \right\rangle =0, \end{align}
 \end{widetext}
 were, $s=1,2,3 \ldots n$. Eq.~\ref{inverse5a} provides $n$ linear equations  for $n$ unknown matrix elements $\langle m|\textbf{M}(\mu)^{-1}|j\rangle$ with $m=1,2 \ldots n$ and therefore can be uniquely determined which in turn helps to determine the rest of the matrix elements 
$\langle i|\textbf{M}(\mu)^{-1}|j\rangle$ following Eq.~\ref{inverse4a}.
 %can therefore
%The above Eq.~\ref{inverse5a} and Eq.~\ref{inverse4a} play a central role to compute the elements of  $\mathbf{M}(\mu)^{-1}$ as well as bare Green's function $\mathbf{g}$.
\section{Transfer matrix eigenvalues in different regimes, exceptional lines, and points for $n=2$}
\label{sec:app2}
In this section, we discuss the nature of eigenvalues of the transfer matrix $\mathbf{T}(\mu)$ for $n=2$, given in Eq.\ref{2T}, in different regimes, exceptional lines, and points as marked in Fig.~\ref{schematic}. We find the eigenvalues analytically by solving for $F(\theta)=0$ with  $F(\theta)$ defined in Eq.~\ref{ftheta}. The analytical results obtained in this section are summarized in Table.~\ref{table1} and Table.~\ref{table2} (third column). \\
%Eq.~\ref{Ftheta} i.e., $F(\theta)=0$. \\
\vskip 0.05cm
\noindent
\textit{Regime (I) in Fig.~\ref{schematic}:}
The regime (I) of Fig.~\ref{schematic} i.e., below the lower band edge corresponds to $\mu<-2t_1-2t_2$. To check the corresponding eigenvalues of the transfer matrix $\mathbf{T}(\mu)$, we set $\mu=-2 t_1-2 t_2 - \varepsilon$ with $\varepsilon>0$. Note that, $\varepsilon>0$ is introduced to indicate that we are accessing regime (I).  The condition $F(\theta)=0$ then provides, 
\begin{align}
\label{belowbandedge}
-2 t_1 -2 t_2 - \varepsilon=-2 t_1 \cos \theta -2 t_2 \cos 2\theta.
\end{align}
The solution of $\theta$ can be written using Eq.~\ref{belowbandedge} as,
\begin{align}
\label{belowlowerbandedge1}
\theta=\cos^{-1}\Big[-\frac{t_1}{4 t_2} \pm \sqrt{\bigg(1+\frac{t_1}{4 t_2}\bigg)^2+\frac{\varepsilon}{4 t_2}}\Big]. %\\ \nonumber 
%\implies \theta=\cos^{-1}[-\frac{t_1}{4 t_2} \pm (1+\frac{t_1}{4 t_2}) \sqrt{1+\frac{\epsilon}{4 t_2}/(1+\frac{t_1}{4 t_2})^2}] 
\end{align}
For the case when we have negative sign in the argument of Eq.~\ref{belowlowerbandedge1}, the argument inside the $\cos^{-1}$ is always less than $-1$. For the case when we have positive sign in the argument of Eq.~\ref{belowlowerbandedge1}, as $\sqrt{(1+\frac{t_1}{4 t_2})^2+\frac{\varepsilon}{4 t_2}}>1+\frac{t_1}{4 t_2}$, thus the argument inside $\cos^{-1}$ is always greater than $1$. As a result, in this below lower band edge regime, the
argument inside  $\cos^{-1}$ is either greater than $1$ or less than $-1$. Therefore, the allowed solutions for $\theta$ is of the form $\theta=c+ i d$ where $c$ can either be $0$(mod $2\pi$) or $\pi$ and $d\in \text{real}$. Thus, the solution $\theta$ does not match with any wave-vector $k$ value for the lattice. As a consequence, all the transfer matrix eigenvalues, given by $e^{i \theta}$, are real with an absolute value not equal to $1$.  \\

%\mathbf{STOP}
\vskip 0.05cm
\noindent
\textit{Regime (II) in Fig.~\ref{schematic}:}
In a similar way, let us consider a small number $\varepsilon>0$ to check the transfer matrix eigenvalues in regime (II) of Fig.~\ref{schematic} where $-2 t_1-2 t_2<\mu<2 t_1-2 t_2$. For this case, we set $\mu=-2 t_1-2 t_2 + \varepsilon$ with $0<\varepsilon< 4 t_1$. The condition $F(\theta)=0$ provides, 
\begin{align}
\label{insidebandedge}
-2 t_1 -2 t_2 + \varepsilon=-2 t_1 \cos \theta -2 t_2 \cos 2\theta.
\end{align}
Using Eq.~\ref{insidebandedge}, we can write the solution of $\theta$ as,
\begin{align}
\label{insidebandedge1}
\theta=\cos^{-1}\Big[-\frac{t_1}{4 t_2} \pm \sqrt{\bigg(1+\frac{t_1}{4 t_2}\bigg)^2-\frac{\varepsilon}{4 t_2}}\Big].
\end{align}
The second term in the argument of Eq.~\ref{insidebandedge1} is always positive in the regime $0<\varepsilon< 4 t_1$  and is bounded as $\varepsilon$ by,
\begin{align}
\label{cases1}
\frac{t_1}{4 t_2}-1 &<\sqrt{\bigg(1+\frac{t_1}{4 t_2}\bigg)^2-\frac{\varepsilon}{4 t_2}}<1+\frac{t_1}{4 t_2}, 
   ~\text{if }~ t_2 < t_1/4 \nonumber\\
0&<\sqrt{\bigg(1+\frac{t_1}{4 t_2}\bigg)^2-\frac{\varepsilon}{4 t_2}}<2,  ~ \text{if}~ t_2 =t_1/4 \nonumber \\
1-\frac{t_1}{4 t_2}&<\sqrt{\bigg(1+\frac{t_1}{4 t_2}\bigg)^2-\frac{\varepsilon}{4 t_2}}<1+\frac{t_1}{4 t_2}, ~\text{if}~ t_2 >t_1/4
\end{align}
With Eq.~\ref{cases1}, for the case when we have positive sign in the argument of Eq.~\ref{insidebandedge1}, then the quantity inside $\cos^{-1}$ is bounded between $-1$ to $1$. This  leads to one real solution  $\theta$ which matches with wave-vector $k$ of the lattice.  In a similar way, for the case when we have negative sign in the argument of Eq.~\ref{insidebandedge1}, then the quantity inside $\cos^{-1}$ is always less than $-1$. This leads  complex solution of $\theta$ of the form $\theta=c+id$ where $c=\pi$ and $d\in \textrm{real}$. As a result, transfer matrix will have two real eigenvalues and two complex conjugate pairs. \\
\vskip 0.05cm
\noindent
\textit{Regime (III) in Fig.~\ref{schematic}:}
Now, to explain the nature of the eigenvalues of transfer matrix $\mathbf{T}(\mu)$ in region (III) in Fig.~\ref{schematic} which is between $2 t_1 - 2 t_2<\mu<\frac{t_1^2}{4 t_2}+ 2 t_2$ with $t_2>t_1/4$, 
 we set $\mu=2 t_1 - 2t_2+ \varepsilon$ with $0<\varepsilon<\varepsilon_c$. At the transition point given by,
 \begin{align}
 \label{epsilonc1}
 \varepsilon_c=4 t_2 \bigg(1-\frac{t_1}{4 t_2}\bigg)^2,
 \end{align}
the chemical potential $\mu$ corresponds to the upper band edge line D of Fig.~\ref{schematic}. With that, $F(\theta)=0$ provides,
\begin{align}
\label{inside band-edge1}
2 t_1 -2 t_2 + \varepsilon = -2 t_1 \cos \theta - 2 t_2 \cos 2 \theta .  
\end{align}
From Eq.~\ref{inside band-edge1}, we can easily write the solution for $\theta$ as,
\begin{align}
\label{inside band-edge1a}
\theta=\cos^{-1}\Big[-\frac{t_1}{4 t_2} \pm \sqrt{\bigg(1-\frac{t_1}{4 t_2}\bigg)^2-\frac{\varepsilon}{4 t_2}}\Big].
\end{align}
 Thus, the second term in Eq.~\ref{inside band-edge1a} is bounded as,
 \begin{align}
 \label{case2}
 0<\sqrt{\bigg(1-\frac{t_1}{4 t_2}\bigg)^2-\frac{\varepsilon}{4 t_2}}<1-\frac{t_1}{4 t_2}.
 \end{align}
 With Eq.~\ref{case2} and since $t_1/4t_2<1$, following Eq.~\ref{inside band-edge1a} for both the cases when we have positive and negative sign in the argument, the entire quantity inside the argument in $\cos^{-1}$ is bounded between $-1$ to $1$. Thus,
 $\theta$ will have two real solutions which matches with wave-vector $k$ of the lattice. Thus transfer matrix eigenvalues will have two complex conjugate pairs. \\
\vskip 0.05cm
\noindent
\textit{Regime (IV) in Fig.~\ref{schematic}:}
Now, to analyse regime (IV) of Fig.~\ref{schematic} i.e. $\mu>\frac{t_1^2}{4 t_2}+ 2 t_2$, we set $\mu=2 t_1-2 t_2+\varepsilon_c+\varepsilon$ with $\varepsilon>0$ and recall that $\varepsilon_c$ is defined in Eq.~\ref{epsilonc1}. The solutions for  $F(\theta)=0$  gives,
\begin{align}
\label{regime4-1}
 2 t_1-2 t_2+ \varepsilon_c+ \varepsilon =-2 t_1 \cos \theta -2 t_2 \cos 2\theta .
\end{align}
Using Eq.~\ref{regime4-1}, we can write the solution for $\theta$ as,
\begin{align}
\label{out band-edge1a}
\theta=\cos^{-1}\bigg[-\frac{t_1}{4 t_2} \pm \sqrt{\bigg(1-\frac{t_1}{4 t_2}\bigg)^2-\bigg(\frac{\varepsilon_c+\varepsilon}{4 t_2}\bigg)}\bigg].  
\end{align}
Using the value of $\varepsilon_c$ [Eq.~\ref{epsilonc1}], Eq.~\ref{out band-edge1a} can be simplified to,
\begin{align}
\label{out band-edge1aa}
\theta =\cos^{-1}\bigg[-\frac{t_1}{4 t_2}\pm i
\sqrt{\frac{\varepsilon}{4 t_2}}\bigg].
\end{align}
Thus, for any $\varepsilon>0$, the solutions of $\theta$ are complex numbers of the form $\theta=c+ i d$ with $c,d\in \mathrm{real}$. This leads to complex solutions of transfer matrix eigenvalues with absolute value never equals to $1$.  \\
\vskip 0.05cm
\noindent
\textit{Regime (V) in Fig.~\ref{schematic}:}
To understand the transfer matrix eigenvalues in regime (V) i.e., $2 t_1 - 2t_2<\mu<2t_2+ \frac{t_1^2}{4 t_2}$ of Fig.~\ref{schematic} with $t_2<t_1/4$. We therefore set, $\mu=2 t_1 - 2t_2+ \varepsilon$ with $0<\varepsilon<\varepsilon_c$. At the value  $\varepsilon_c$ (Eq.~\ref{epsilonc1}) $\mu$ hits the exceptional line E of Fig.~\ref{schematic}. Then the solution $F(\theta)=0$ gives,
\begin{align}
\label{above upper band-edge}
2 t_1 -2 t_2 + \varepsilon = -2 t_1 \cos \theta - 2 t_2 \cos 2 \theta .  
\end{align}
Using Eq.~\ref{above upper band-edge}, the solution for $\theta$ can be written as,
\begin{align}
\label{above upper band edge 1a}
\theta=\cos^{-1}\Big[-\frac{t_1}{4 t_2} \pm \sqrt{\bigg(\frac{t_1}{4 t_2}-1\bigg)^2-\frac{\varepsilon}{4 t_2}}\Big].   
\end{align}
 Since $t_1/4t_2>1$, the second term of Eq.~\ref{above upper band edge 1a} is bounded as,
  \begin{align}
 \label{case3}
 0<\sqrt{\bigg(\frac{t_1}{4 t_2}-1\bigg)^2-\frac{\varepsilon}{4 t_2}}<\frac{t_1}{4 t_2}-1.
 \end{align}
 From Eq.~\ref{case3}, for both the cases when we have positive and negative sign in the argument of Eq.~\ref{above upper band edge 1a}, then the entire argument in $\cos^{-1}$ is less than $-1$. Thus,
the solutions of $\theta$  have the form $\theta=c+id$ with $c=\pi$ and $d\in \mathrm{real}$ and therefore these eigenvalues do not match with wave-vector $k$ of the lattice. Thus all the eigenvalues of transfer matrix $\mathbf{T}(\mu)$ are real with absolute value not equal to $1$. \\
\vskip 0.05cm
\noindent
\textit{Exceptional line A in Fig.~\ref{schematic}:}
 When chemical potential $\mu$ is at the lower band edge $\mu=-2t_1-2 t_2$ along line A of Fig.~\ref{schematic}, $F(\theta)=0$ gives,
\begin{align}
\label{lower bandedge1}
    -2 t_1- 2 t_2=-2 t_1 \cos \theta - 2 t_2 \cos 2\theta. 
\end{align}
Eq.~\ref{lower bandedge1} can be simplified to,
\begin{align}
\label{app_lower band edge}
4 \sin^2 \frac{\theta}{2} \bigg(t_1 + 4 t_2 \cos^2 \frac{\theta}{2}\bigg)=0.
\end{align}
Thus, the solutions for $\theta$'s are 
\begin{align}
\label{solution}
\theta=0,~\cos^{-1}\bigg[\!\!-\!\!\bigg(1+ \frac{t_1}{2 t_2}\bigg)\bigg].
\end{align}
As transfer matrix eigenvalues are $e^{i \theta}$, $\theta=0$ will give  two eigenvalues as $1$. Thus, we immediately see that any point corresponding to the lower band edge (line A of Fig.~\ref{schematic}) is always an exceptional point of underlying transfer matrix. Now, as the ratio of $t_1/t_2$ is always positive, the argument in $\cos^{-1}$ is always less than $-1$. Thus, the other solution for $\theta$ has form $\theta=c+id$ with $c=\pi$ and $d\in \mathrm{real}$. Thus, at the lower band edge, transfer matrix has exceptional point with two eigenvalues $1$ and  two other eigenvalues are real with absolute value not equal to $1$.\\
\vskip 0.05cm
\noindent
\textit{Exceptional line B in Fig.~\ref{schematic}}
When chemical potential $\mu$ at the upper band edge i.e. $\mu=2t_1-2t_2$  with $t_2<t_1/4$ along line B in Fig.~\ref{schematic}, $F(\theta)=0$  gives, 
\begin{align}
\label{upper band-edge}
  2 t_1- 2 t_2=-2 t_1 \cos \theta - 2 t_2 \cos 2\theta.
\end{align}
Eq.~\ref{upper band-edge} can be simplified to,
\begin{align}
\label{upper band-edge1}
4 \cos^2 \frac{\theta}{2} \bigg(\!\!-t_1 + 4 t_2 \sin^2 \frac{\theta}{2}\bigg)=0.
\end{align}
Thus, the solutions for $\theta$'s are 
\begin{align}
\label{upperc4}
\theta= \pi ,  \cos^{-1}\bigg[1-\frac{t_1}{2t_2}\bigg].
\end{align}
As transfer matrix eigenvalues are $e^{i \theta}$, $\theta= \pi$  will give the two eigenvalues as $-1$. Thus, once again we  immediately see the upper band edge also corresponds to transfer matrix exceptional point. Now, since $t_1/t_2>4$ , the other  solution of $\theta$ has the form $\theta=c+id$ with $c=\pi, d\in \mathrm{real}$. Thus, along line B of Fig.~\ref{schematic}, transfer matrix has exceptional point with two eigenvalue $-1$ and  two other eigenvalues are real numbers with absolute value not equal to $1$.\\
\vskip 0.05cm
\noindent
\textit{Exceptional line C in Fig.~\ref{schematic}:}
\label{C}
When the chemical potential $\mu$ is along the line C of Fig.~\ref{schematic} i.e. $\mu=\frac{t_1^2}{4 t_2}+ 2 t_2$ with $t_2>t_1/4$, it corresponds to the upper band edge with wave-vector $k\neq \pi$.  Along this line, $F(\theta)=0$ gives,
\begin{align}
\label{linecupper}
 \frac{t_1^2}{4 t_2}+ 2 t_2 =-2 t_1 \cos \theta -2 t_2 \cos 2\theta.
 \end{align}
 Using Eq.~\ref{linecupper}, the solution of $\theta$'s are ,
 \begin{align}
 \label{lineupperc1}    
\theta =\cos^{-1}\bigg[-\frac{t_1}{4 t_2}\bigg].
\end{align}
Since the transfer matrix eigenvalues are $e^{\pm i \theta}$, using Eq.~\ref{lineupperc1} we can write the eigenvalues as,
\begin{align}
\label{lineupperc2}
-\frac{t_1}{4 t_2} + i \sqrt{1-\bigg(\frac{t_1}{4 t_2}\bigg)^2},~~ -\frac{t_1}{4 t_2} - i \sqrt{1-\bigg(\frac{t_1}{4 t_2}\bigg)^2}.
\end{align} 
Since $t_1/4t_2 <1$, these eigenvalues are complex with absolute value $1$. Thus, the upper band edge along line C of the Fig.~\ref{schematic} has two pairs of complex exceptional point as mentioned it Eq.~\ref{lineupperc2}.\\
\vskip 0.05cm
\noindent
 \textit{Exceptional line D in Fig.~\ref{schematic}:}
When the chemical potential $\mu=2 t_1 - 2 t_2$ with $t_2 >t_1/4$ along line D of Fig.~\ref{schematic}, from Eq.~\ref{upper band-edge} and Eq.~\ref{upper band-edge1}, in a similar way, the solutions of $\theta$ are,
\begin{align}
 \theta=\pi, ~ \cos^{-1}\bigg[1-\frac{t_1}{2t_2}\bigg]. 
\end{align}
Thus two transfer matrix eigenvalues are $-1$ (exceptional points) along this line D. Since, $0<t_1/t_2<4$, the other  solution for $\theta$ is bounded between $-1$ to $1$. Thus, the other two eigenvalues of the transfer matrix are complex conjugate pairs with an absolute value $1$. \\
\vskip 0.05cm
\noindent
\textit{Exceptional line E in Fig.~\ref{schematic}:}
To understand the transfer matrix eigenvalues along line E  i.e. $\mu=\frac{t_1^2}{4 t_2}+ 2 t_2$ of Fig.~\ref{schematic} with $t_2>t_1/4$, we have to follow the same analysis as done for the case of along exceptional line C. Thus, eigenvalues of transfer matrix are,
\begin{align}
-\frac{t_1}{4 t_2} + \sqrt{\Big(\frac{t_1}{4 t_2}\Big)^2-1},~~ -\frac{t_1}{4 t_2} - \sqrt{\Big(\frac{t_1}{4 t_2}\Big)^2-1}. 
\end{align}
Since, $t_1/4t_2>1$, all the eigenvalues are real with absolute value not equal to $1$.\\
\vskip 0.05cm
\noindent
\textit{Exceptional point $\Gamma_e$ in Fig.~\ref{schematic}:}
When the chemical potential $\mu$ is at the upper band edge i.e. $\mu=2t_1-2 t_2$ with $t_1/4t_2=1$ (at $\Gamma_e$ point of Fig.~\ref{schematic}), $F(\theta)=0$ gives,, 
\begin{align}
\label{lower band-edgeaa}
4 \cos^2 \frac{\theta}{2} \bigg(-t_1 + 4 t_2 \sin^2 \frac{\theta}{2}\bigg)=0.
\end{align}
This is exactly same as Eq.~\ref{upper band-edge1} with $t_2=t_1/4$. Thus, exactly like Eq.~\ref{upperc4} the solutions for $\theta$'s are 
\begin{align}
\theta= \pi,~~\cos^{-1}\bigg[1-\frac{t_1}{2t_2}\bigg]. 
\end{align}
Since, $t_2=t_1/4$, all the transfer matrix eigenvalues are $-1$. Thus, at this point, transfer matrix has fourth order exceptional point. 

We have given the analytical results of transfer matrix eigenvalues for all the cases in Table.~\ref{table1} and .~\ref{table2}. Also specifically, we have shown the plot for transfer matrix eigenvalues in Fig.~\ref{eigenvalues1}. 

\bibliography{ref_finite_range_hopping}

\end{document}